\documentclass[conference]{IEEEtran}
\IEEEoverridecommandlockouts
\usepackage{cite}
\usepackage{amsmath,amssymb,amsfonts}
\usepackage{algorithm}
\usepackage{subfig}
\usepackage{amsfonts}
\usepackage{amssymb}
\usepackage{gensymb}
\usepackage{algorithmic}
\usepackage{graphicx}
\usepackage{textcomp}
\usepackage{xcolor}
\def\BibTeX{{\rm B\kern-.05em{\sc i\kern-.025em b}\kern-.08em
    T\kern-.1667em\lower.7ex\hbox{E}\kern-.125emX}}
\begin{document}

\title {Widely-distributed Radar Imaging Based on Consensus ADMM\\
\thanks{This work is supported by Luxembourg National Research Fund (FNR) through the CORE project “SPRINGER” under Grant C18/IS/12734677/SPRINGER.}}
\author{\IEEEauthorblockN{Ruizhi Hu, Bhavani Shankar Mysore Rama Rao, Ahmed Murtada,\\ Mohammad Alaee-Kerahroodi, and Bj\"orn Ottersten}
\IEEEauthorblockA{\textit{Interdisciplinary Centre for Security, Reliability and Trust (SnT)} \\
\textit{University of Luxembourg}\\
Luxembourg City, Luxembourg \\
\{ruizhi.hu, bhavani.shankar, ahmed.murtada, mohammad.alaee, bjorn.ottersten\}@uni.lu}}
\maketitle
\begin{abstract}
A widely-distributed radar system is a promising architecture to enhance imaging performance. However, most existing algorithms rely on isotropic scattering assumption, which is only satisfied in colocated radar systems. Moreover, due to noise and imaging model imperfections, artifacts such as layovers are common in radar images. In this paper, a novel $l_1$-regularized, consensus alternating direction method of multipliers (CADMM) based algorithm is proposed to mitigate artifacts by exploiting the spatial diversity in a widely-distributed radar system. By imposing the consensus constraints on the local images formed by the distributed antenna clusters and solving the resulting distributed optimization problem, the scenario's spatially-invariant common features are retained and the spatially-variant artifacts are mitigated in a data-driven fashion. The iterative procedure will finally will finally converge to a high-quality global image in the consensus of all widely-distributed measurements. The proposed algorithm outperforms the existing joint sparsity-based composite imaging (JSC) algorithm in terms of artifacts mitigation. It can also reduce the computation and storage burden of large-scale imaging problems through its distributed and parallelizable optimization scheme.
\end{abstract}

\begin{IEEEkeywords}
ADMM, artifacts mitigation, consensus, distributed optimization, distributed radar imaging
\end{IEEEkeywords}

\section{Introduction}
Advances in technology and signal processing have furthered the proliferation of privacy-preserving, all-weather, and all-day radar systems for imaging. Applications include security, assisted living, smart-buildings, and health monitoring \cite{gurbuz2019radar,gennarelli2019radar}, to name a few. In many of these applications, to avoid occlusions and achieve high angular resolution, widely-distributed radar systems are anticipated to have better performance by exploiting the spatial diversity \cite{haimovich2007mimo}. However, for imaging applications on such systems, the conventional isotropic scattering of targets is no longer satisfied. Thus novel imaging algorithms need to be explored.

There have been a limited number of studies on widely-distributed radar imaging in the literature. Two relevant imaging genres are
the extension of classical radar imaging with a cluster of colocated antennas and the wide-angle synthetic aperture radar (WASAR) imaging. For the former genre, optimization-based imaging algorithms are proposed to obtain the image and resolve position ambiguity \cite{mansour2018radar} and clock ambiguity \cite{lodhi2019coherent} simultaneously of a system with a clusters of colocated antennas. However, in both works, the
antennas are deployed such that the isotropic scattering can be assumed. Hence, this technique cannot be applied in a truly widely-distributed system, where targets exhibit angle-dependent scattering.
On the other hand, there are two algorithmic approaches in WASAR.
The first one is based on the attributed scattering center model, which characterizes the canonical scattering behavior of scatterers with parameters \cite{potter1997attributed}. Nevertheless, the imaging involves a dictionary-based learning process that is computationally intensive \cite{hammond2013sar,yang2019robust}. The second is the composite imaging algorithm \cite{moses2004wide,sanders2017composite,hu2017video}. It divides the whole synthetic aperture into sub-apertures that are more in line with the isotropic scattering assumption and implements image formation for each sub-aperture. Then the sub-aperture images are fused by pixel-wise maximization. Composite imaging is sub-optimal because the information from multiple aspects is not fully exploited; clearly, the fusion process can be improved by utilizing this additional information.

Consensus alternating direction method of multipliers (CADMM) is a general framework for distributed optimization \cite{boyd2011distributed}, it has been applied to radar signal processing, such as detection \cite{zabolotsky2018distributed}, waveform design \cite{wang2019design}, and compressive antenna imaging \cite{heredia2017norm}. To overcome the drawbacks and inability of the existing solutions \cite{sanders2017composite,mansour2018radar,lodhi2019coherent} to exploit opportunities offered by widely-distributed radar imaging systems, a novel algorithm based on the CADMM framework is proposed in this paper. Contrary to the joint sparsity-based composite (JSC) imaging algorithm in \cite{sanders2017composite} that fuses the sub-aperture/local images after imaging, in the proposed algorithm, information from multiple antenna clusters are fused in an iterative data-driven way during the optimization-based imaging process. By imposing the additional consensus constraints, each local image will be consistent with the corresponding measurements from each antenna cluster and, at the same time, gradually converge to a global image. In this way, the spatially-invariant common features are kept while the spatially-variant artifacts like layover effects are mitigated. Due to the distributed nature of the CADMM framework, parallelization can be implemented to enhance its efficiency. 
% The proposed algorithm can also be extended to similar radar imaging architectures such as WASAR and MIMO radar.
\section{Signal Model and Joint Sparsity-based Composite Imaging}
\label{sec:intro}
\label{sec:intro}
\begin{figure}[!htbp]
	\centering
	\includegraphics[width=2.8 in]{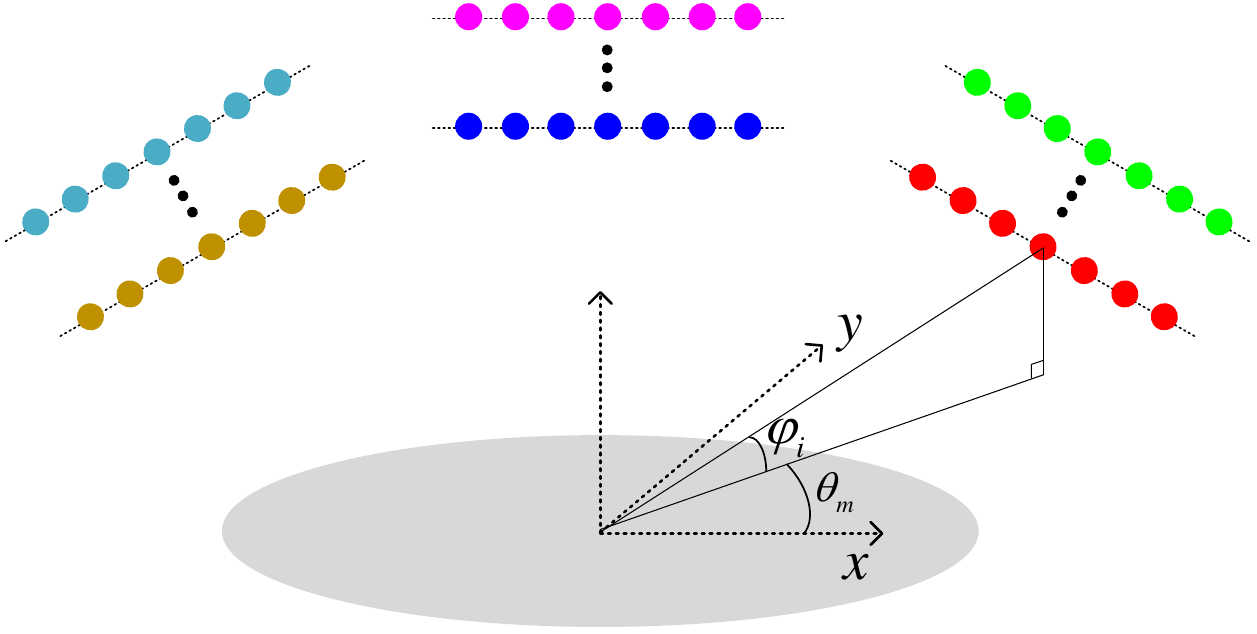}
	\caption{Geometry of distributed radar system.}
	\label{fig_geo}
\end{figure}
The system geometry is illustrated in Fig.~\ref{fig_geo}, where the dots of the same color indicate a particular antenna phase centers (APC) cluster. 
% The APCs could be formed by either a real aperture or a synthetic aperture. 
We further assume a distributed mono-static architecture that a cluster will not process signals from other clusters. In this architecture, the different clusters operate independently without interference due to orthogonal resources (time slots, frequency bands, etc). 

% the other clusters do not receive the echo during the working slot of a specific cluster. Although this configuration does not make full use of the spatial diversity provided by a widely-distributed radar system and has a longer data acquisition time, it has advantages in terms of simpler model and signal processing. Nevertheless, through proper signal separation and adaption of the bi-static range model, the proposed algorithm can also be extended to more advanced architectures like multiple-input-multiple-output (MIMO) radar.

When isotropic scattering is assumed, the image can be obtained by 2D (non-uniform) Fourier transform \cite{munson1983tomographic,hu2019refocusing,hu2020orthorectified}, or by solving an ill-posed inverse problem with prior information such as sparsity or smoothness \cite{ccetin2014sparsity,sanders2017composite,kantor2019polar}. However, because of the large azimuth and elevation angular extent in a widely-distributed radar system, isotropic scattering cannot be satisfied across all antennas. Instead, we impose isotropic scattering only within each APC cluster $i \in \left\{ {1,2,...,I} \right\}$. The $i$-th cluster has a constant elevation angle $\varphi _i$ and a limited extent of azimuth angles $\left\{ \theta _m \right\} _{m=m_i}^{M_i}$, where $m_i$ and $M_i$ denote the starting and ending indices and $N_i=M_i-m_i+1$ is the number of azimuth samples in $i-$th cluster. 
Under 2D tomographic radar imaging framework \cite{munson1983tomographic,sanders2019fourier}, after dechirping, the signal of the antenna at $\theta _m$ in $i-$th cluster is given by
\begin{equation}
\label{eq:radon}
y^i\left( k,m \right) =\iint\limits_{\Omega}{\tilde{r}^i\left( x,y \right) e^{j\omega _{k}^{i}\left( x\cos \theta _m+y\sin \theta _m \right)}dxdy},
\end{equation}
where $\tilde{r}^i\left( x,y \right)$  denotes the corresponding complex scattering coefficient of a ground target at $(x,y)$ for the $i-$th cluster. $\Omega$ is the ground footprint, which is assumed to be the same for all antennas. $\left\{ \omega _{k}^{i}=4\pi f_k\cos \varphi _i/c \right\} _{k=1}^{K}$ denotes the wavenumber under ${\varphi_i}$, ${f_k}$ denotes the linear beat frequency, and $K$ is the number of fast-time samples. 
By discretizing the imaging scene into $L$ uniformly distributed pixels $\left\{ \left( x_l,y_l \right) \right\} _{l=1}^{L}$ and stacking $\left\{ {{y^i}\left( {k,m} \right)} \right\}_{k = 1}^K$ into a vector ${\mathbf{y}}_m^i$, the matrix form of (\ref{eq:radon}) is given by
\begin{equation}
\label{eq:matrix}
\begin{gathered}
\mathbf{y}_{m}^{i}=\mathbf{A}_{m}^{i}\mathbf{\tilde{r}}^i+\mathbf{n}_{m}^{i}\in \mathbb{C}^{K\times 1},
\hfill \\
\mathbf{A}_{m}^{i}=\left[ \boldsymbol{\alpha }_{m}^{1},\boldsymbol{\alpha }_{m}^{2},\cdots ,\boldsymbol{\alpha }_{m}^{L} \right] \in \mathbb{C}^{K\times L},
\hfill \\
\boldsymbol{\alpha }_{m}^{l}=\left[ \begin{array}{c}
	e^{j\omega _{1}^{i}\left( x_l\cos \theta _m+y_l\sin \theta _m \right)}\\
	e^{j\omega _{2}^{i}\left( x_l\cos \theta _m+y_l\sin \theta _m \right)}\\
	\vdots\\
	e^{j\omega _{K}^{i}\left( x_l\cos \theta _m+y_l\sin \theta _m \right)}\\
\end{array} \right] \in \mathbb{C}^{K\times 1},
\hfill \\
\end{gathered} 
\end{equation}
where ${\mathbf{A}}_m^i$ is the measurement matrix for ${\left( {{\theta _m},{\varphi _i}} \right)}$, $\mathbf{n}_{m}^{i}$ denote the error from model imperfections, noise, measurement error, etc.

Based on (\ref{eq:matrix}), the measurement process of each antenna can be explicitly implemented by matrix multiplication of $\mathbf{A}_m^i$, or alternatively, be implemented as an operator via 1D non-uniform FFT \cite{sanders2017composite,sanders2019fourier} according to the projection-slice theorem \cite{munson1983tomographic}. 
Through a further vertical stacking of $\left\{ \mathbf{y}_{m}^{i} \right\} _{m=m_i}^{M_i}\rightarrow \mathbf{y}^i$ and $\left\{ \mathbf{A}_{m}^{i} \right\} _{m=m_i}^{M_i}\rightarrow \mathbf{A}^i$, the whole measurement process for the $i$-th APC cluster is given by
\begin{equation}
\label{eq:measure_i}
\mathbf{y}^i=\mathbf{A}^i\mathbf{\tilde{r}}^i+\mathbf{n}^i\in \mathbb{C}^{N_iK\times 1}
\end{equation}

Besides the assumptions on the measurement model, there are other factors that will cause artifacts in the image, such as synchronization/measurement error, occlusion, shadow, noise, and a dynamic scenario. In particular, for targets located outside of the ground plane, their reconstructed locations will deviate from their actual locations, which is the well-known layover effect \cite{brenner2008radar,hu2020orthorectified}. The layover targets may overlay on the ground targets, resulting in artifacts that are undesired for post-processing. Additionally, the number of unknowns (number of pixels) is usually larger than the number of measurements. All these factors render the measurement model in (\ref{eq:measure_i}) ill-posed, and a reasonable solution can only be obtained from prior knowledge through regularization.

One effective solution is using joint sparsity-based composite (JSC) imaging to mitigate the anisotropic scattering \cite{sanders2017composite}, which comprises two steps. The first step is to independently solve $l_1$-regularized optimization problems for each antenna cluster to obtain $I$ magnitude images $\left\{ {\mathbf{r}}^{i} \right\} _{i=1}^{I}$ via
\begin{equation}
\label{eq:jsc}
{\mathbf{r}}^{i}=\mathrm{arg}\mathop {\min} \limits_{{\mathbf{r}}^i}\left\{ \frac{\mathrm{\mu}}{2}\left\| \mathbf{y}^i-\mathbf{A}^i\mathbf{\Theta }^i{\mathbf{r}}^i \right\| _{2}^{2}+\left\| {\mathbf{r}}^i \right\| _1 \right\},
\end{equation}
where $\mu >0$ is a hyperparameter to trade-off the data fidelity term and the $l_1$-regularizer.
$\mathbf{\Theta }^i\in \mathbb{C}^{L\times L}$ is a diagonal unitary matrix whose entries are the estimated phase angles of the pixels in $\mathbf{r}^i$, i.e. $\mathbf{\tilde{r}}^i\approx \mathbf{\Theta }^i\mathbf{r}^i
$. The value of $\mathbf{\Theta }^i$ can be estimated from some initial approximate solutions \cite{ccetin2012handling,sanders2017composite}, such as from back-projection  $\mathbf{\Theta }^i=\mathrm{diag}\left\{ \exp \left[ j\angle \left( \mathbf{A}^{iH}\mathbf{y}^i \right) \right] \right\}$, where $\mathbf{A}^{iH}\in \mathbb{C}^{L\times N_iK}$ is the conjugate transpose of $\mathbf{A}^{i}$ and $\angle$ is the angle of a complex number. 
 
The optimization in (\ref{eq:jsc}) can be solved by proximal gradient algorithms \cite{parikh2014proximal}, such as the fast iterative shrinkage-thresholding algorithm (FISTA) \cite{beck2009fast}, or in the ADMM manner \cite{sanders2017composite}. The final JSC image $\mathbf{r}_{J}$ is obtained via a pixel-wise maximization of the $I$ magnitude images $\left\{ {\mathbf{r}}^{i} \right\} _{i=1}^{I}$ as
\begin{equation}
\mathbf{r}_J\left( l \right) =\max \left\{ {\mathbf{r}}^{1}\left( l \right) ,{\mathbf{r}}^{2}\left( l \right) ,...,{\mathbf{r}}^{I}\left( l \right) \right\} 
\end{equation}
\section{Consensus-ADMM Imaging}
While JSC imaging is useful in solving artifacts caused by anisotropic scattering, some artifacts such as layover still remain evident as they are relatively strong, and the pixel-wise maximization cannot remove them. Although it is acknowledged that the range profiles and phase of the scene vary significantly from different views, we could try to find a global magnitude image that fits the measurements from all APC clusters by employing the consensus ADMM (CADMM) framework. Using the terminology from CADMM, the sub-aperture images are henceforth referred to as local images. Due to the imposed consensus constraints on local images, the common features among local images are retained while the non-shared artifacts are diminished during the optimization process of the CADMM imaging algorithm. Through the proposed imaging algorithm, the measurements from widely-distributed antenna clusters can collaborate in the data domain to converge to an artifacts-mitigated global image in a data-driven fashion.   
The conversion to CADMM imaging algorithm is straightforward by adding the consensus constraints to (\ref{eq:jsc}), then the optimization problem becomes
\begin{equation}
    \label{eq:cadmm_problem}
\begin{gathered}
\underset{{\mathbf{r}}^1,{\mathbf{r}}^2,\cdots ,{\mathbf{r}}^I}{\min}\sum_{i=1}^I{\left\{ \frac{\mathrm{\mu}}{2}\left\| \mathbf{y}^i-\mathbf{A}^i\mathbf{\Theta }^i
{\mathbf{r}}^i \right\| _{2}^{2}+\left\| {\mathbf{r}}^i \right\| _1 \right\}}
\\
\,\,\,\,\,\,\,\,\,\,\,\,\,\,\,\,\,\,s.t.\,\,\,\,\,\,\,\,{\mathbf{r}}^i=\mathbf{g},\,\,\,\,\,\,\,i=1,2,...,I.
\end{gathered}
\end{equation}
where $\mathbf{g}$ denotes the global magnitude image.

Let ${\mathbf{r}}=\left\{ {\mathbf{r}}^i \right\} _{i=1}^{I}$ and $\boldsymbol{\sigma }=\left\{ \boldsymbol{\sigma }^i \right\} _{i=1}^{I}$, the augmented Lagrangian of (\ref{eq:cadmm_problem}) can be written as a sum of $I$ terms,
\begin{equation}
\label{eq:aug_lag}
\begin{gathered}
 \mathcal{L}\left( {\mathbf{r}},\mathbf{g},\boldsymbol{\sigma } \right) =\sum_{i=1}^I{\mathcal{L}_i\left( {\mathbf{r}}^i,\mathbf{g},\boldsymbol{\sigma }^i \right)}, \hfill \\
\mathcal{L}_i\left( {\mathbf{r}}^i,\mathbf{g},\boldsymbol{\sigma }^i \right) 
= \left\{ \begin{gathered}
\frac{\mu}{2}\left\| \mathbf{y}^i-\mathbf{A}^i\mathbf{\Theta }^i{\mathbf{r}}^i \right\| _{2}^{2}+\left\| \mathbf{g} \right\| _1 + ... \hfill \\
+\left. \langle \boldsymbol{\sigma }^i,{\mathbf{r}}^i-\mathbf{g} \right. \rangle +\frac{\beta}{2}\left\| {\mathbf{r}}^i-\mathbf{g} \right\| _{2}^{2} \hfill \\ 
\end{gathered}  \right\}, \hfill \\
\end{gathered} 
\end{equation}
where $\left< \cdot \right>$  denotes the inner product of vectors, ${{\boldsymbol{\sigma }}^i}$ is the dual variable, and $\beta >0$ is the augmented Lagrangian penalty hyperparameter.

It is noted that in (\ref{eq:aug_lag}), $\left\| \mathbf{{r}}^i \right\| _1$ is replaced by $\left\| \mathbf{g} \right\| _1$, which corresponds to a centralized model where the major optimization is implemented on the global image. Instead, if $\left\| \mathbf{{r}}^i \right\| _1$ is used, the major optimizations are implemented on local images, which follows the common CADMM framework \cite{boyd2011distributed} and corresponds to a decentralized model. For a comparison of the two models, we defer it to a later work. The imposed consensus constraints are critical in our algorithm, from which the global image $\mathbf{g}$ is obtained adaptively and jointly from the widely-distributed radar measurements. Similar to the implementation of ADMM \cite{boyd2011distributed}, CADMM is implemented via alternating updates to ${\mathbf{r}}$, $\mathbf{g}$, and $\boldsymbol{\sigma}$. The resulting algorithm is shown in Algorithm \ref{alg_cadmm} and the detailed explanations are given in the sequel.
\begin{algorithm}[t]  \caption{\small Solving problem (\ref{eq:cadmm_problem}) by CADMM}  \textbf{Input:} The measurement matrices/operators $\left\{ \mathbf{A}^i \right\} _{i=1}^{I}$ and the measured signal $\left\{ \mathbf{y}^i \right\} _{i=1}^{I}$.\\  
\textbf{Initialize:} The iteration counter $t=0$, the maximum number of iterations ${t_{\max }}$, the hyperparameters $\mu$ and $\beta$, the tolerance $\epsilon$, the local images $\left\{ {\mathbf{r}}_{0}^{i} \right\} _{i=1}^{I}=\left| \mathbf{A}^{iH}\mathbf{y}^i \right|$, the estimated phase angles
$\left\{ \mathbf{\Theta }^i \right\} _{i=1}^{I}=\mathrm{diag}\left\{ \exp \left[ j\angle \left( \mathbf{A}^{iH}\mathbf{y}^i \right) \right] \right\}$, the dual variables $\left\{ \boldsymbol{\sigma }_{0}^{i} \right\} _{i=1}^{I}=\mathbf{0}$, the global image $\mathbf{g}_0=\mathbf{0}$.\\  \textbf{Output:} $\mathbf{g}^*$. \\  
\textbf{while} $t < {t_{\max }}$ or the stopping criterion (\ref{eq:stopping}) not satisfied \textbf{do}  
\begin{enumerate}    
\item Update $\left\{ {\mathbf{r}}_{t+1}^{i} \right\} _{i=1}^{I}$ (possibly in parallel) based on (\ref{eq:update_r}).  
\item Update ${\mathbf{g}}_{t + 1}$ based on (\ref{eq:g1_h}).
\item Update $\left\{ \boldsymbol{\sigma }_{t+1}^{i} \right\} _{i=1}^{I}$ based on (\ref{eq:update_sigma}).
\item $t\gets t+1$.
\end{enumerate}  
\textbf{end while}  
\label{alg_cadmm}  
\end{algorithm}

\subsection{Update of ${\mathbf{r}}$}
\label{update_r}
Let $\mathbf{g}_t$ and $\boldsymbol{\sigma }_{t}$ denote the values of $\mathbf{g}$ and $\boldsymbol{\sigma }$ after the $t-$th iteration. Since $\mathcal{L}\left( {\mathbf{r}},\mathbf{g},\boldsymbol{\sigma } \right)$ is decomposable with respect to ${{\mathbf{r}}^i}$, the updated value of $\mathbf{{r}}_{t+1}$ can be obtained by each of $\mathbf{{r}}_{t+1}^{i}$ in parallel as
\begin{equation}
\begin{gathered}
\mathbf{{r}}_{t+1}^{i}=\mathrm{arg}\mathop {\min} \limits_{{\mathbf{r}}^i}\mathcal{L}_i\left( {\mathbf{r}}^i;\mathbf{g}_t,\boldsymbol{\sigma }_{t}^{i} \right) \hfill \\
=\mathrm{arg}\mathop {\min} \limits_{{\mathbf{r}}^i}\frac{\mu}{2}\left\| \mathbf{y}^i-\mathbf{A}^i\mathbf{\Theta }^i{\mathbf{r}}^i \right\| _{2}^{2}+\left. \langle \boldsymbol{\sigma }_{t}^{i},{\mathbf{r}}^i \right. \rangle +\frac{\beta}{2}\left\| {\mathbf{r}}^i-\mathbf{g}_t \right\| _{2}^{2} \hfill \\ 
\end{gathered}
\end{equation}

Because $\mathcal{L}_i\left( {\mathbf{r}}^i;\mathbf{g}_t,\boldsymbol{\sigma }_{t}^{i} \right)$ is differentiable with respect to ${{{\mathbf{r}}^i}}$, ${\mathbf{r}}_{t + 1}^i$ can be obtained explicitly by letting $\nabla _{{\mathbf{r}}^i}\mathcal{L}_i=\mathbf{0}$, which is equivalent to
\begin{equation}
\label{eq:update_r}
\left( \mu \mathbf{\Theta }^{i*}\mathbf{A}^{iH}\mathbf{A}^i\mathbf{\Theta }^i+\beta \mathbf{I}_L \right) {\mathbf{r}}^i=\mu \mathbf{\Theta }^{i*}\mathbf{A}^{iH}\mathbf{y}^i-\boldsymbol{\sigma }_{t}^{i}+\beta \mathbf{g}_t,
\end{equation}
where $\mathbf{\Theta }^{i*}$ is the conjugate of $\mathbf{\Theta }^i$ and $\mathbf{I}_L$ is the $ L\times L$ identity matrix. The matrix $\left( \mu \mathbf{\Theta }^{i*}\mathbf{A}^{iH}\mathbf{A}^i\mathbf{\Theta }^i+\beta \mathbf{I}_L \right)$ in (\ref{eq:update_r}) is positive definite as $\mu$ and $\beta$ are positive, and hence $\mathbf{{r}}_{t+1}^{i}$ can be solved directly by matrix inversion, or as in our case iteratively by conjugate gradient descent algorithm \cite{shewchuk1994introduction}.

\subsection{Update of ${\mathbf{g}}$}
After obtaining ${\mathbf{r}}_{t+1}=\left\{ {\mathbf{r}}_{t+1}^{i} \right\} _{i=1}^{I}$, the current $\mathbf{g}_{t+1}$ can be obtained via solving another sub-problem of ${\mathbf{g}}$,
\begin{equation}
\label{eq:problem_g}
\begin{gathered}
\mathbf{g}_{t+1}=\mathrm{arg}\mathop {\min} \limits_{\mathbf{g}}\mathcal{L}\left( \mathbf{g};{\mathbf{r}}_{t+1},\boldsymbol{\sigma }_t \right) \hfill \\
=\mathrm{arg}\mathop {\min} \limits_{\mathbf{g}}\sum_{i=1}^I{\left\{ \left\| \mathbf{g} \right\| _1-\left< \boldsymbol{\sigma }_{t}^{i},\mathbf{g} \right> +\frac{\beta}{2}\left\| {\mathbf{r}}_{t+1}^{i}-\mathbf{g} \right\| _{2}^{2} \right\}}
\hfill \\
\end{gathered}
\end{equation}

The objective function above involves information from all $I$ clusters and is not decomposable with respect to $\mathbf{g}$, it also involves a non-differentiable function $\left\| \mathbf{g} \right\|_1$. To solve this sub-problem via proximal gradient algorithms \cite{parikh2014proximal}, the objective function is divided into the non-differentiable part $\left\| \mathbf{g} \right\|_1$ and the differentiable part $h\left( \mathbf{g};{\mathbf{r}}_{t+1},\boldsymbol{\sigma }_t \right)$,
\begin{equation}
h\left( \mathbf{g};\mathbf{{r}}_{t+1},\boldsymbol{\sigma }_t \right) =\frac{1}{I}\sum_{i=1}^I{\left[ \frac{\beta}{2}\left\| \mathbf{{r}}_{t+1}^{i}-\mathbf{g} \right\| _{2}^{2}\left. -\langle \boldsymbol{\sigma }_{t}^{i},\mathbf{g} \right. \rangle \right]}
\end{equation}

Through this decomposition, (\ref{eq:problem_g}) becomes 
\begin{equation}
\label{eq:g1_h}
\mathbf{g}_{t+1}=\mathrm{arg}\mathop {\min} \limits_{\mathbf{g}}\left\{ \left\| \mathbf{g} \right\| _1+h\left( \mathbf{g};{\mathbf{r}}_{t+1},\boldsymbol{\sigma }_t \right) \right\},
\end{equation}
and is solved by FISTA with backtracking \cite{beck2009fast}.
\begin{figure*}[h]
	\centering
	\subfloat[]{\includegraphics[width=1.7 in]{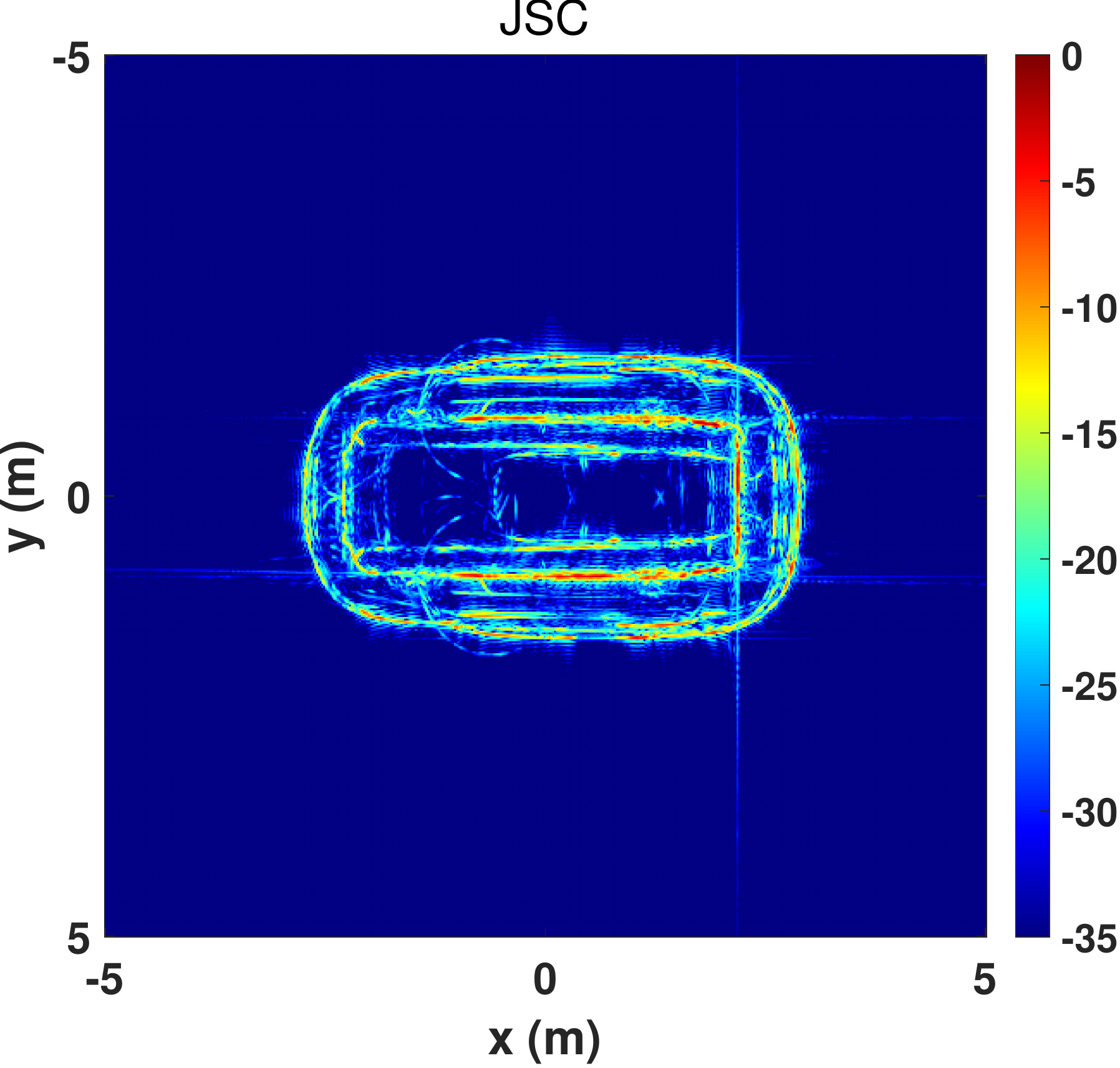}}
	\hfil
	\subfloat[]{\includegraphics[width=1.7 in]{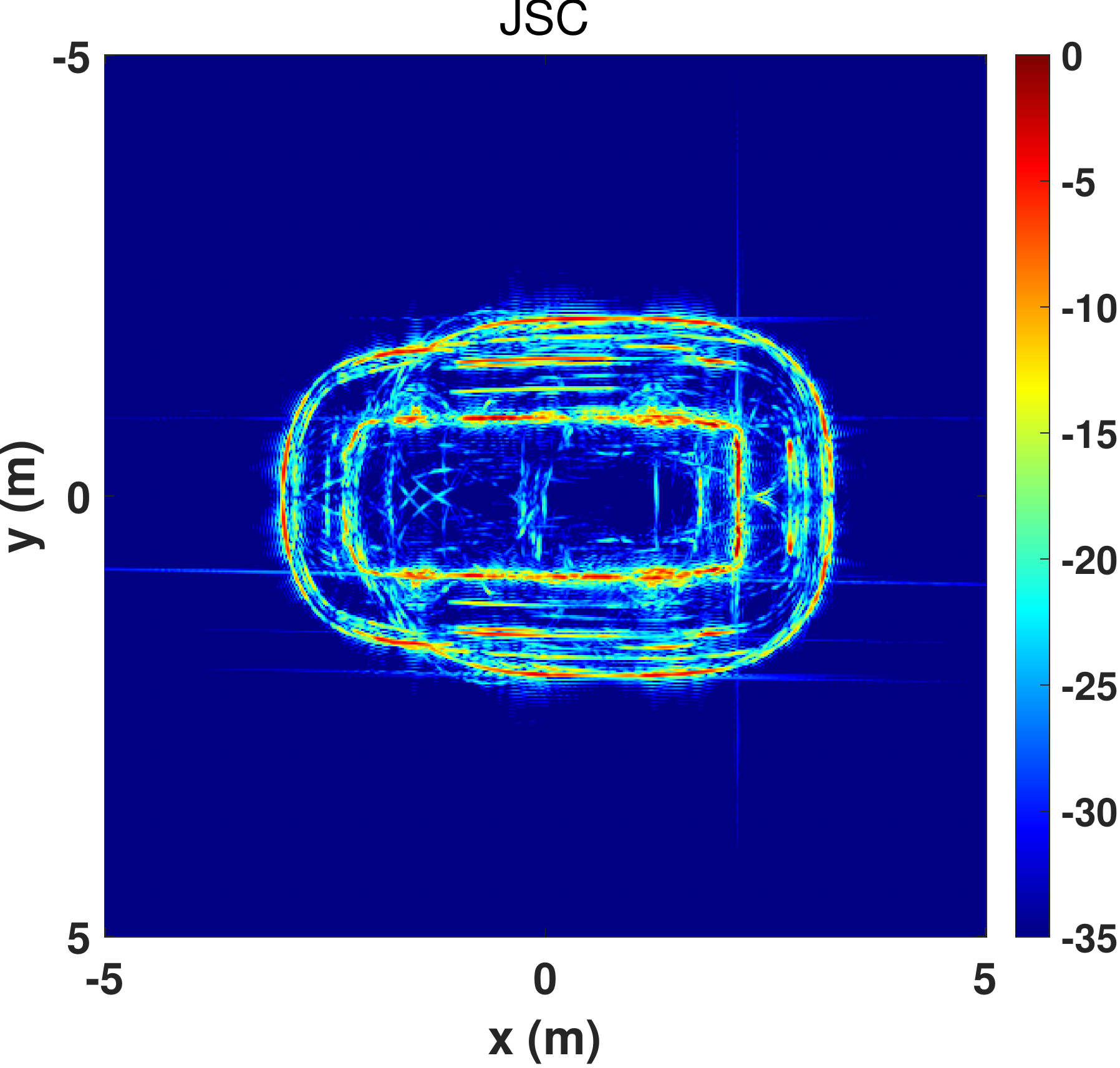}}
	\hfil
	\subfloat[]{\includegraphics[width=1.7 in]{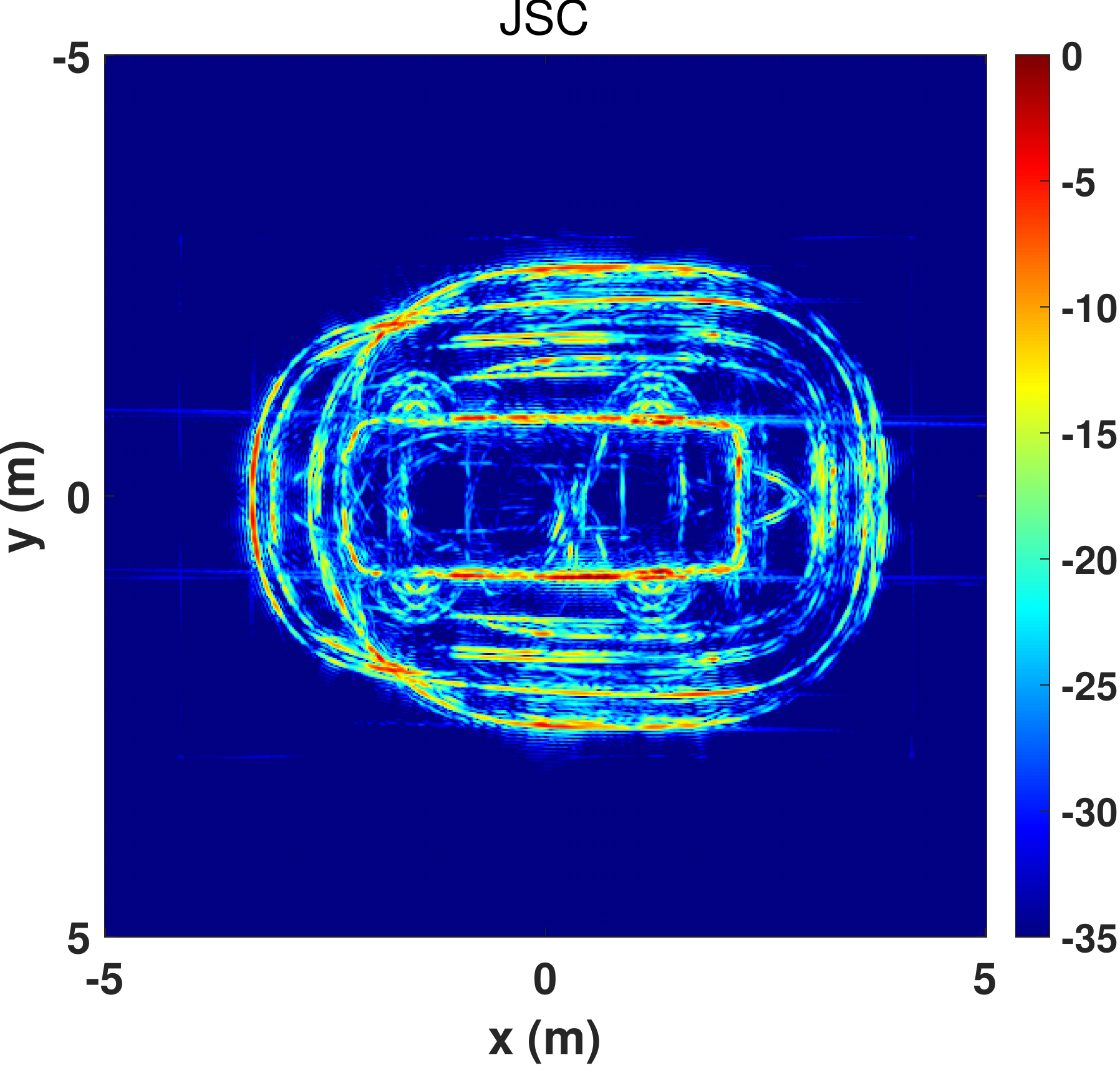}}
	\hfil
	\subfloat[]{\includegraphics[width=1.7 in]{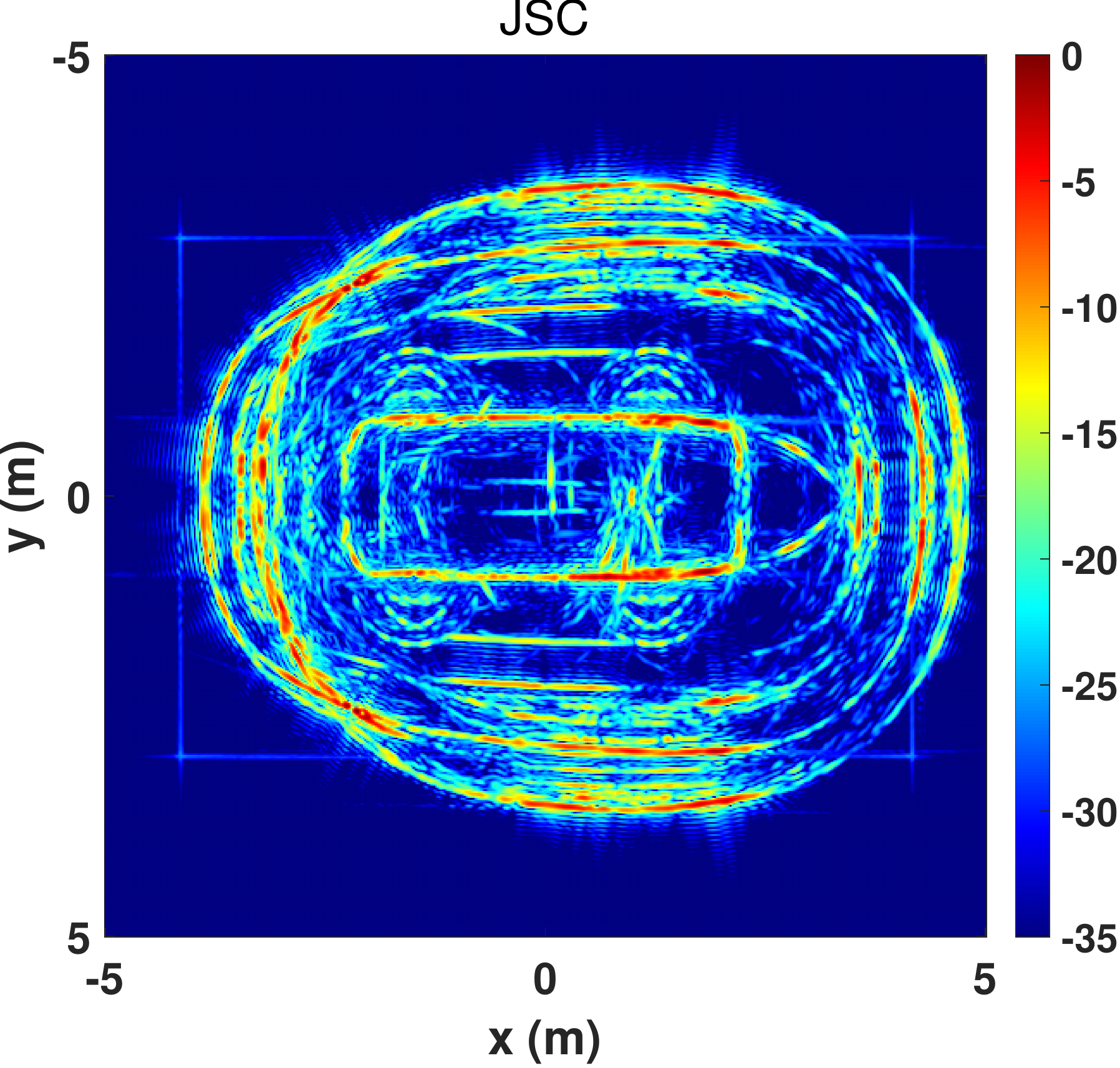}}
    \hfil
	\subfloat[]{\includegraphics[width=1.7 in]{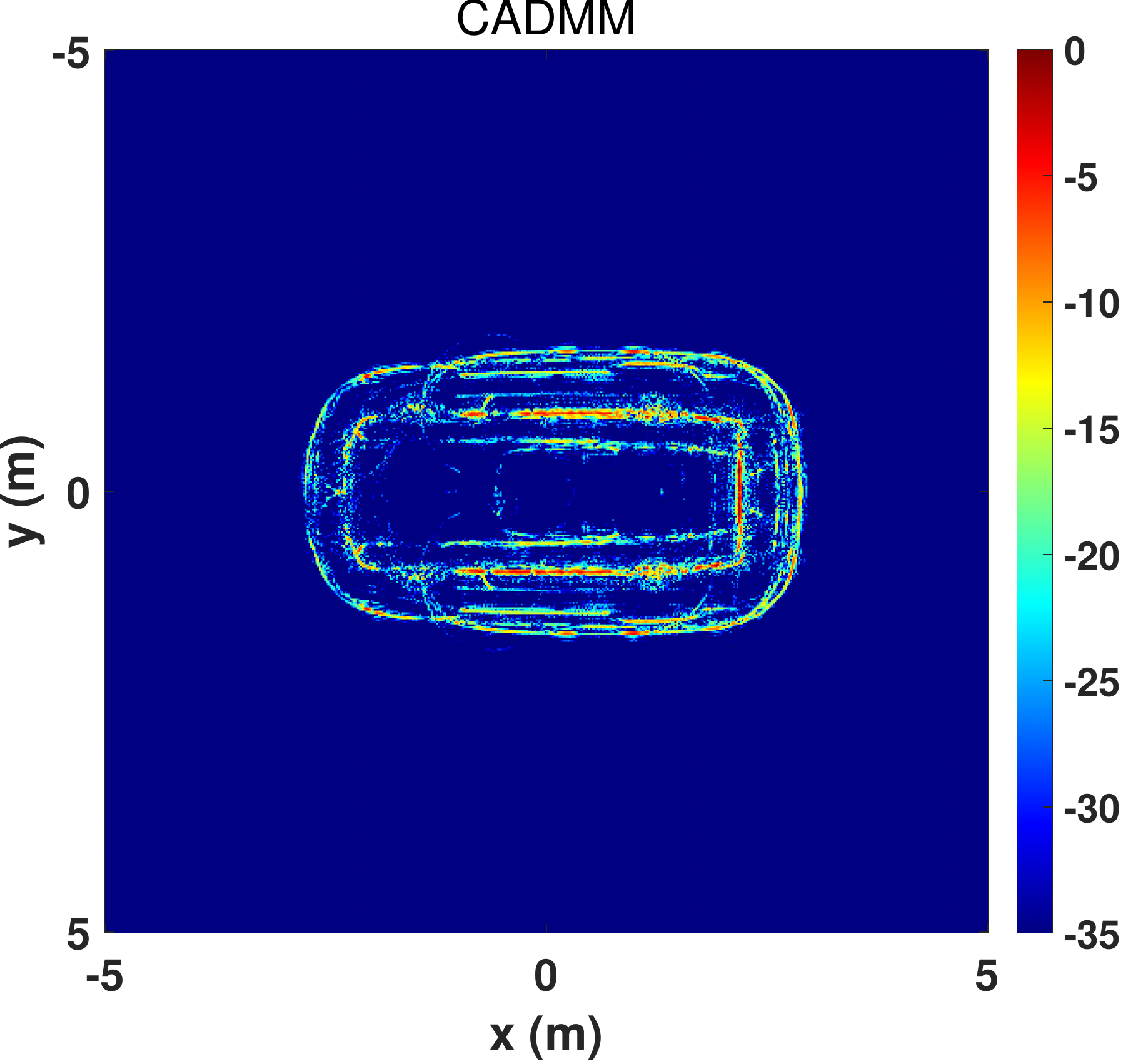}}
	\hfil
	\subfloat[]{\includegraphics[width=1.7 in]{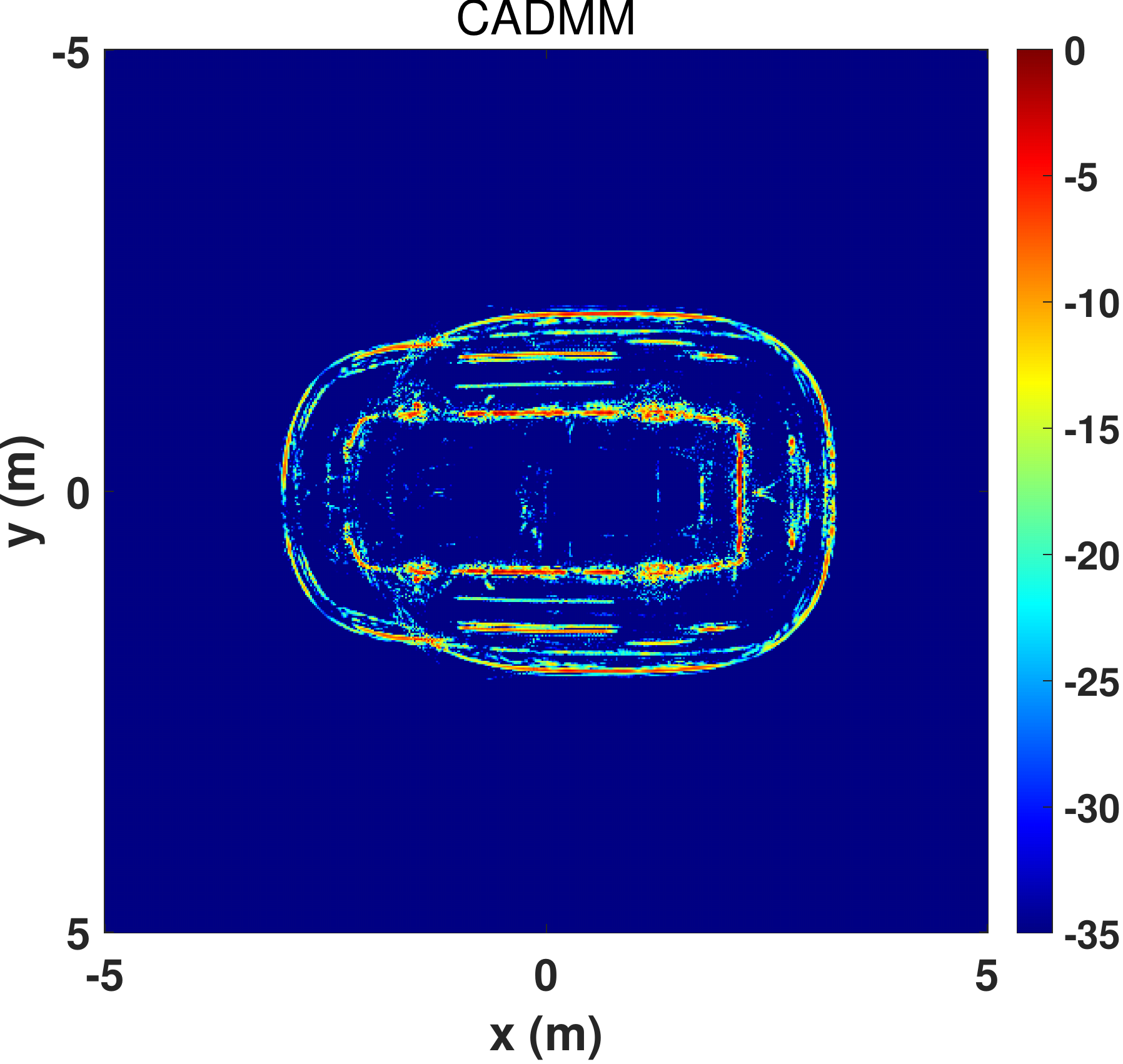}}
	\hfil
	\subfloat[]{\includegraphics[width=1.7 in]{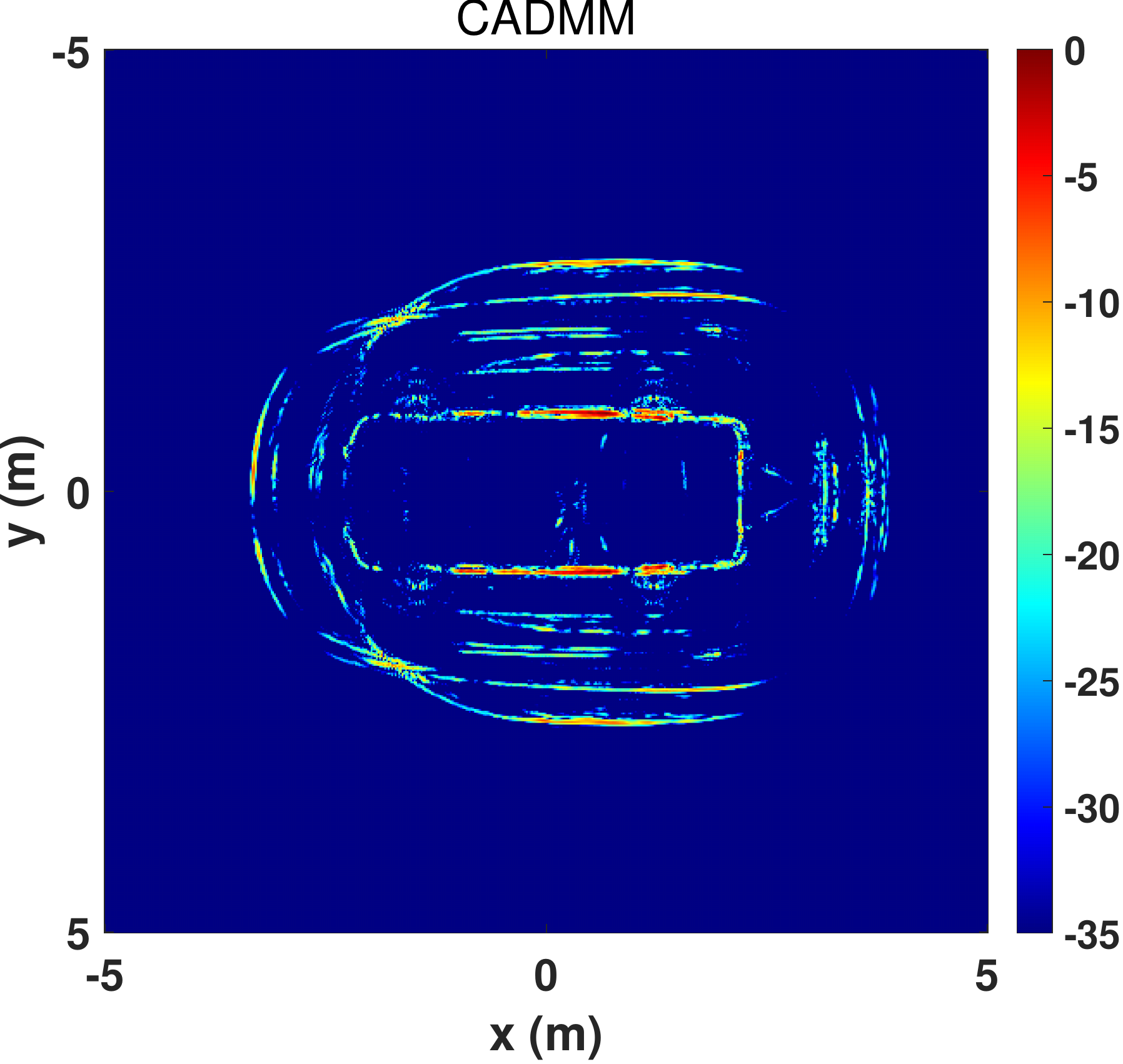}}
	\hfil
	\subfloat[]{\includegraphics[width=1.7 in]{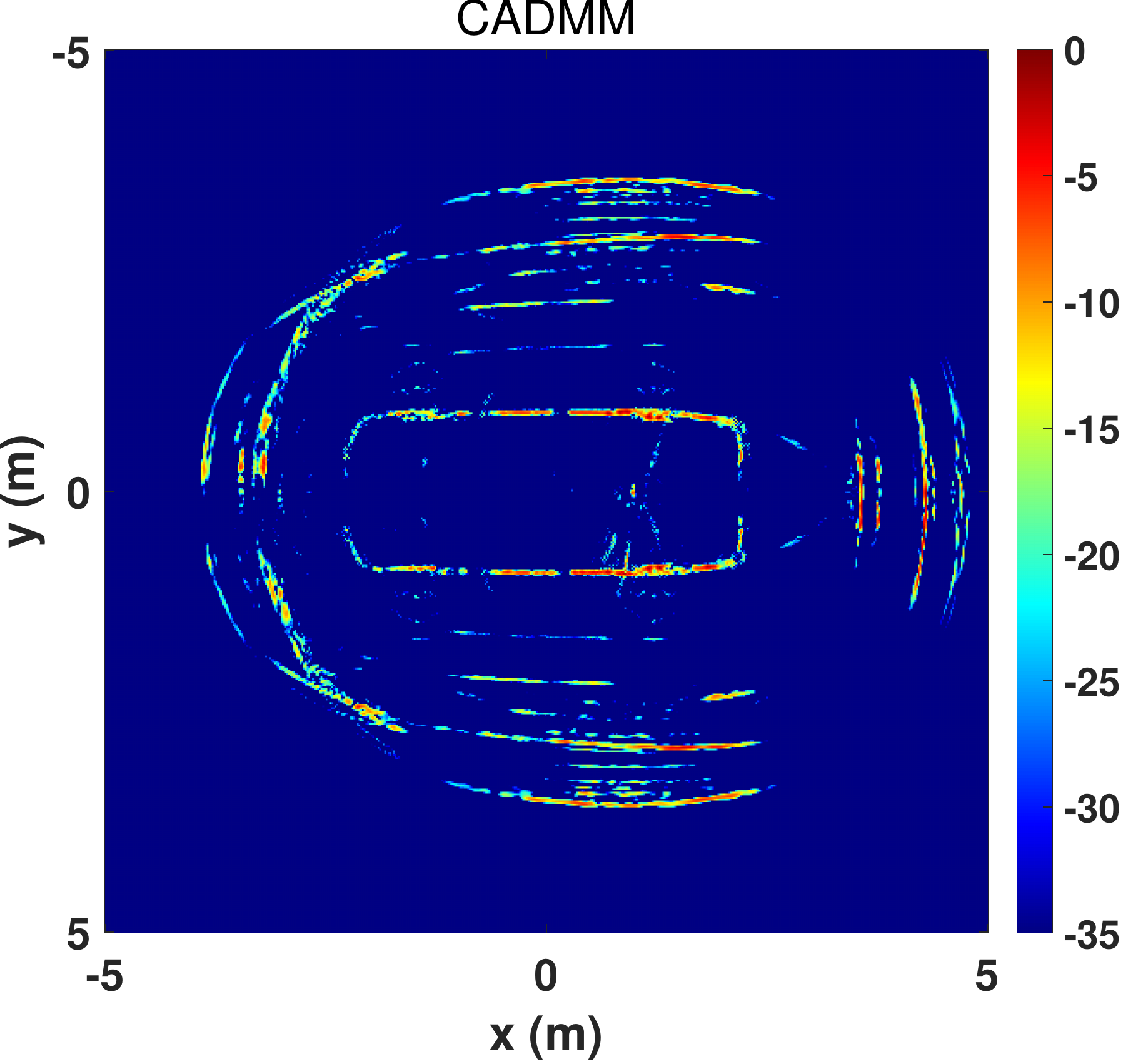}}
	\caption{(a)-(d) JSC images with (a) $\varphi _i=30^{\degree}$; (b) $\varphi _i=40^{\degree}$; (c) $\varphi _i=50^{\degree}$; (d) $\varphi _i=60^{\degree}$. (e)-(h) CADMM images with (e) $\varphi _i=30^{\degree}$; (f) $\varphi _i=40^{\degree}$; (g) $\varphi _i=50^{\degree}$; (h) $\varphi _i=60^{\degree}$.}
	\label{fig_360}
\end{figure*}
\subsection{Update of $\boldsymbol{\sigma }$ and stopping criterion}
Given the current value of ${\mathbf{r}}_{t+1}$ and
$\mathbf{g}_{t+1}$, the final step of each iteration is updating the dual variable via
\begin{equation}
\label{eq:update_sigma}
\boldsymbol{\sigma }_{t+1}^{i}=\boldsymbol{\sigma }_{t}^{i}+\beta \left( {\mathbf{r}}_{t+1}^{i}-\mathbf{g}_{t+1} \right) 
\end{equation}

The aforementioned updates to ${\mathbf{r}}$, $\mathbf{g}$, and $\boldsymbol{\sigma }$ are repeated until convergence \cite{boyd2011distributed}. We define $\boldsymbol{\eta }_{t}^{pri}=\sum_{i=1}^I{\left( {\mathbf{r}}_{t}^{i}-\mathbf{g}_t \right)}$ as the primal residual, and $\boldsymbol{\eta }_{t}^{dual}=\mathbf{g}_t-\mathbf{g}_{t-1}$ as the dual residual after the $t-$th iteration, respectively, and
$\epsilon$ as the absolute tolerance.
The stopping criterion is given by
\begin{equation}
\label{eq:stopping}
\left\| \boldsymbol{\eta }_{t}^{pri} \right\| _2 < \epsilon \,\,\, \text{and} \,\,\,\left\| \boldsymbol{\eta }_{t}^{dual} \right\| _2 < \epsilon. 
\end{equation}
\section{Experiments and discussion}
To validate the effectiveness of the proposed CADMM imaging algorithm, the CV domes sample dataset \cite{dungan2010civilian} is utilized in our experiments. It is a simulated data of ten civilian vehicle facet models. For each model, a X-band electromagnetic simulation produced fully polarized, far-field, mono-static scattering for   a $360{\degree}$ azimuth angles, and elevation angles from $30{\degree}$ to $60{\degree}$. The online sample dataset provides only ${N_\varphi } = 4$ elevation angles $\varphi _i=\left\{ 30^{\degree},40^{\degree},50^{\degree},60^{\degree} \right\}$. By dividing the elevation/azimuth angles into multiple clusters, the data can be regarded as a simulation of a widely-distributed mono-static radar system. 
\begin{figure}[!htbp]
	\centering
	\subfloat[]{\includegraphics[width=1.74 in]{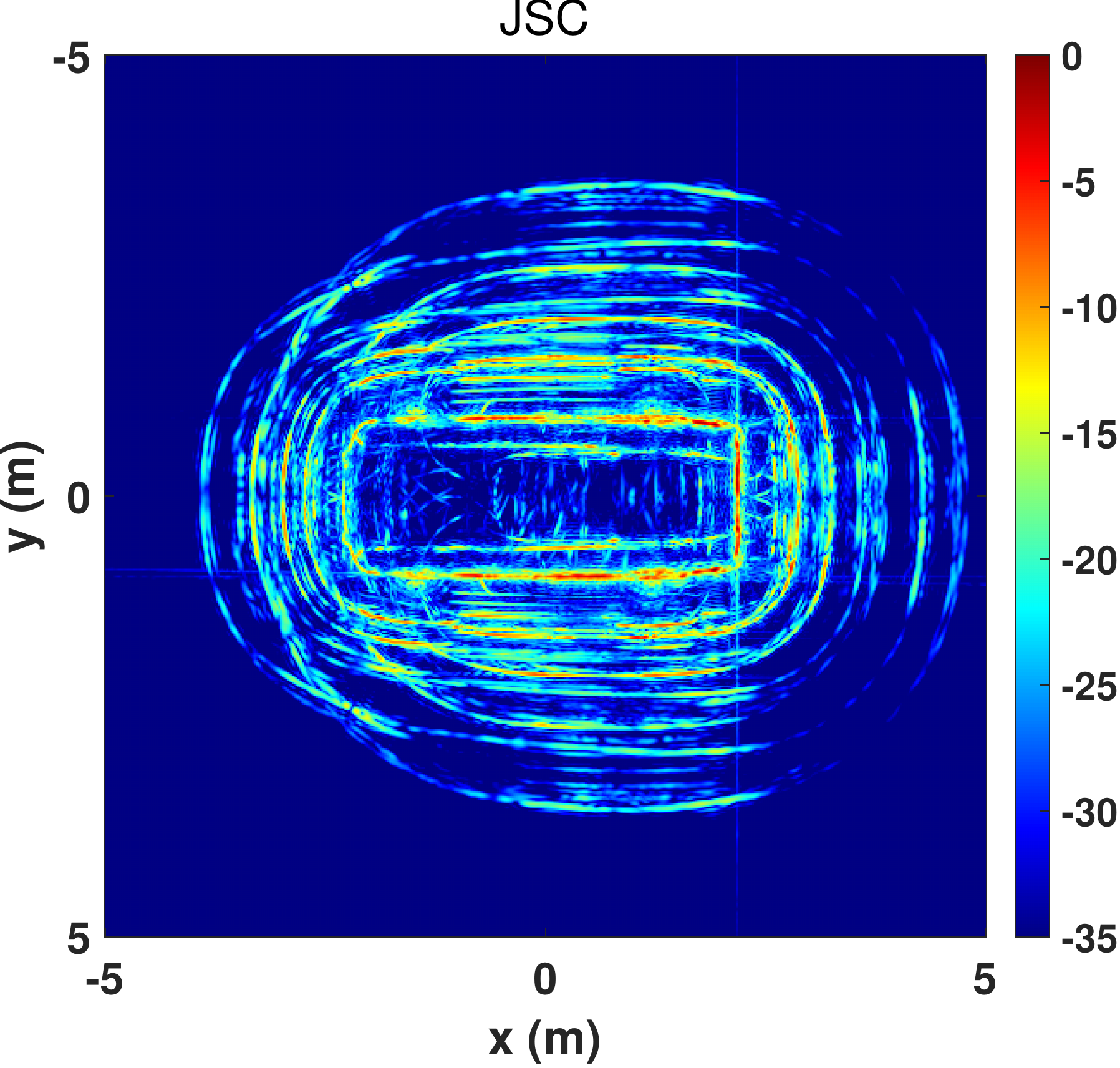}}
	\hfil
	\subfloat[]{\includegraphics[width=1.74 in]{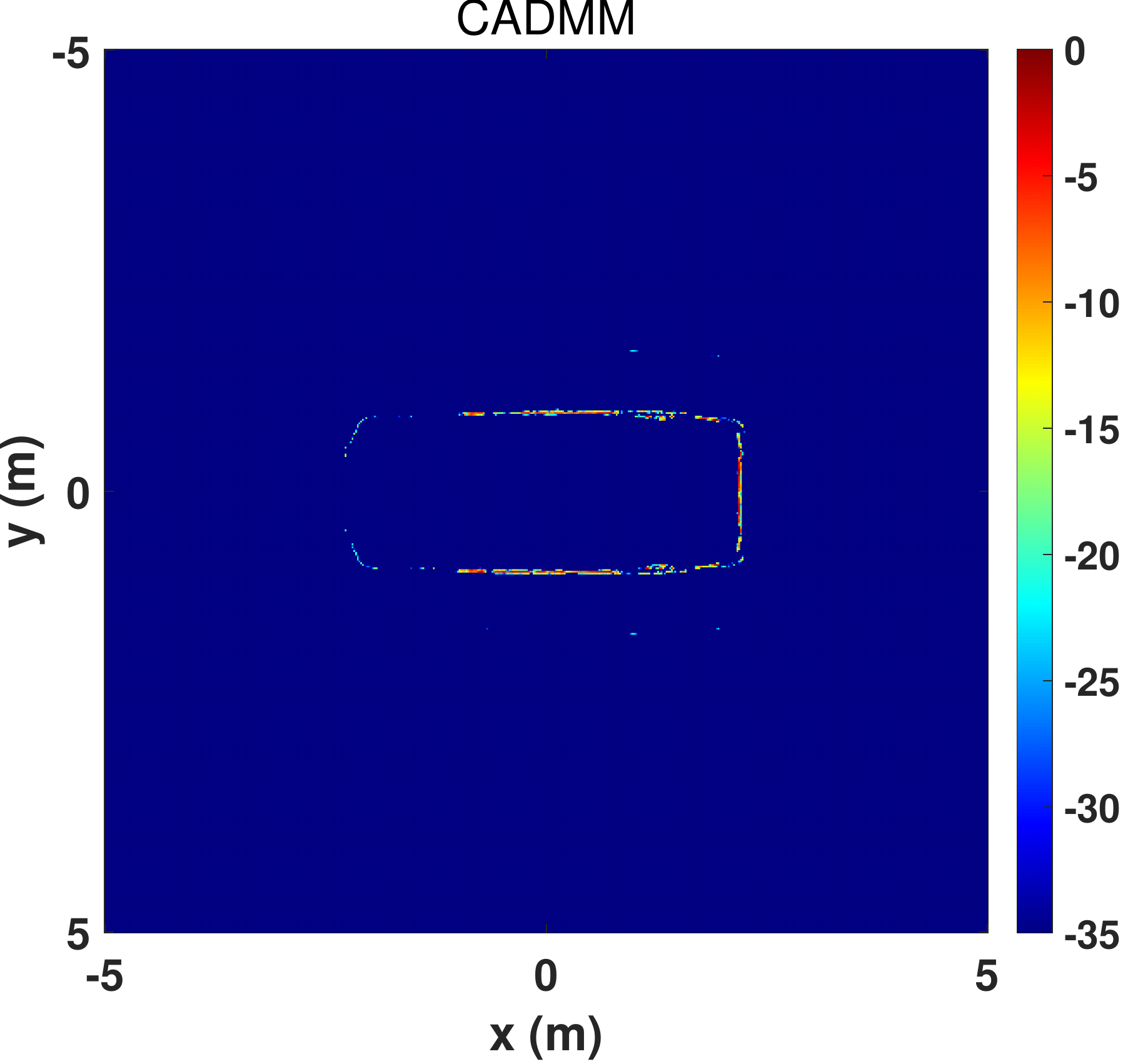}}
	\hfil
	\subfloat[]{\includegraphics[width=1.74 in]{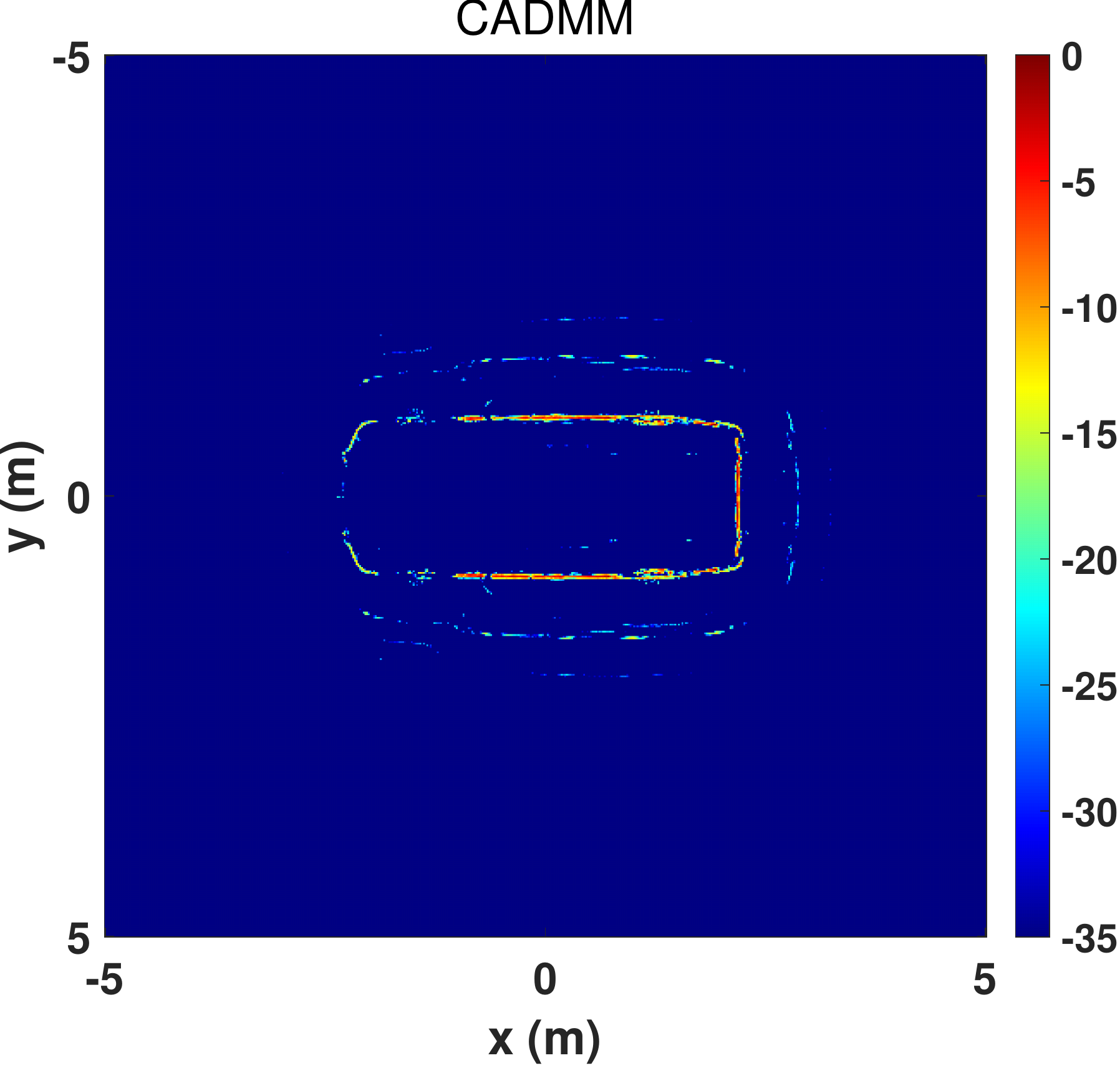}}
	\hfil
	\subfloat[]{\includegraphics[width=1.74 in]{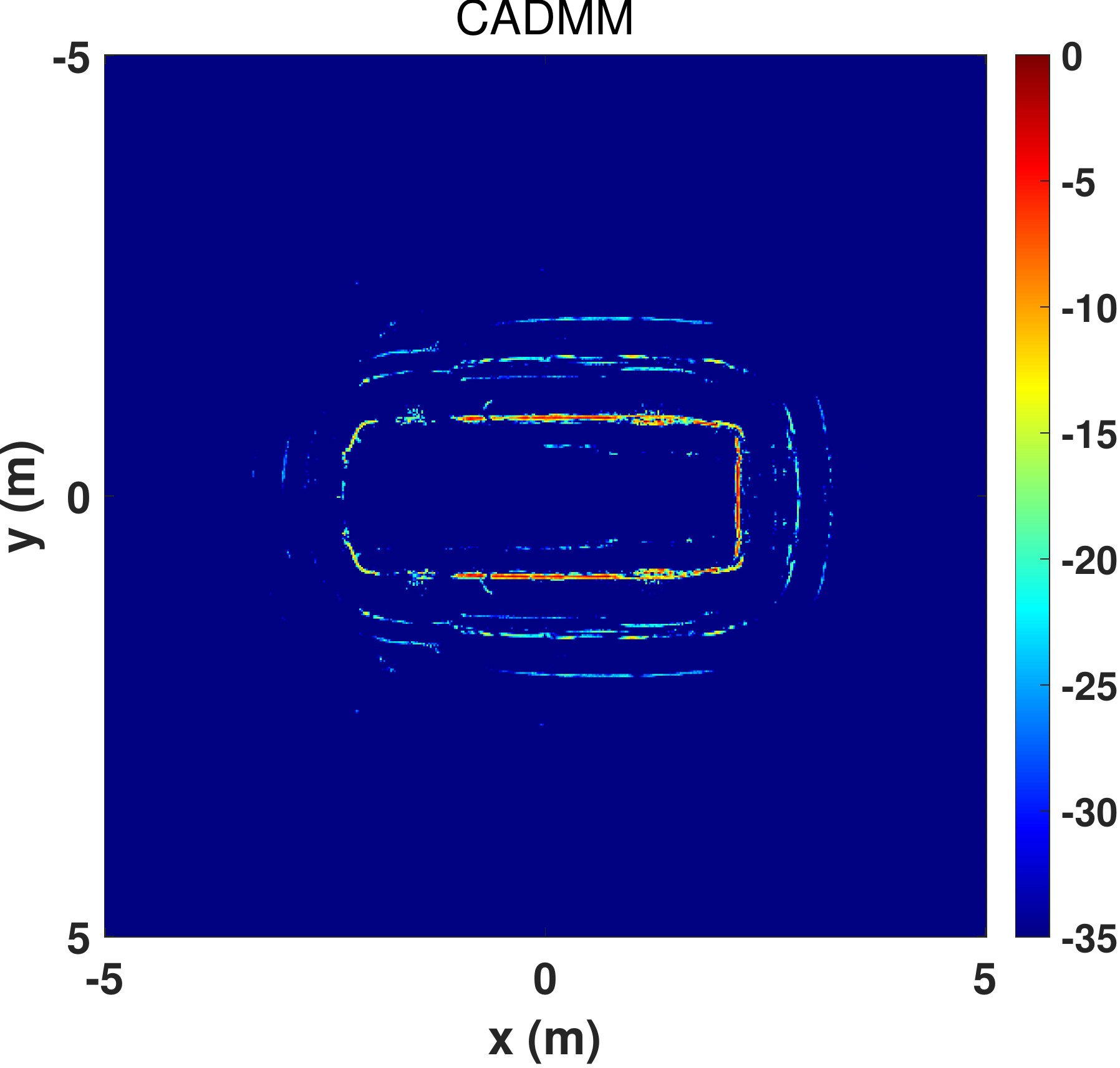}}
	\caption{Full data results. (a) JSC; (b) CADMM ($\mu =100$); (c) CADMM ($\mu =200$); (d) CADMM ($\mu =300$).}
	\label{fig_full}
\end{figure}
\subsection{Results from full $360^{\degree}$ data}
Firstly, to show the superior imaging quality of the proposed algorithm, the full $360^{\degree}$ data of `Jeep99' model with HH polarization is utilized, and the JSC algorithm is implemented as a comparison. For both algorithms, the $360^{\degree}$ data for each $\varphi _i$ is equally divided into ${N_\theta } = 20$ APC clusters without overlapping, and each cluster has $18^{\degree}$ in azimuth. All imaging results are shown for magnitude above -35 dB. Except as otherwise noted, $\mu =100$ is set for both JSC and CADMM,  $\beta =50$ and $\epsilon = {\text{0.01}}$ are set for CADMM.

The results of both algorithms are shown in Fig.~\ref{fig_360}, where Fig.~\ref{fig_360}(a-d) are the JSC images $\left( \varphi _i=30^{\degree}, 40^{\degree}, 50^{\degree}, 60^{\degree} \right)$ and Fig.~\ref{fig_360}(e-f) are the corresponding CADMM images. For each $\varphi _i$ with the same cluster division in azimuth, CADMM presents better images than JSC in terms of lower side-lobe and more concentrated energy due to the additional consensus constraints. For both algorithms, the outer parts of images expand with the increasing elevation angles, demonstrating the spatially-variant layover artifacts. On the contrary, the rectangles on the image centers remain approximately the same for different $\varphi _i$, which could be inferred as the true target on the ground. 

To mitigate the layover artifacts by further exploiting the variance along elevation, data from the 4 elevation angles are all used, while the same azimuth division $\left( 20\times 18^{\degree} \right)$ is kept. The results are given in Fig.~\ref{fig_full}, where Fig.~\ref{fig_full}(a) is the JSC image through a pixel-wise maximization across all sub-images of the 4 elevation angles. The image quality of Fig.~\ref{fig_full}(a) became even worse because the layover artifacts are strong as shown in Fig.~\ref{fig_360}(a-d), so they are preserved by the pixel-wise maximization. As a contrast, in Fig.~\ref{fig_full}(b), layover artifacts are almost eliminated because of the additional spatial diversity from different $\varphi _i$ exploited by CADMM. 

\begin{figure}[htbp]
\centerline{\includegraphics[width=3.2 in]{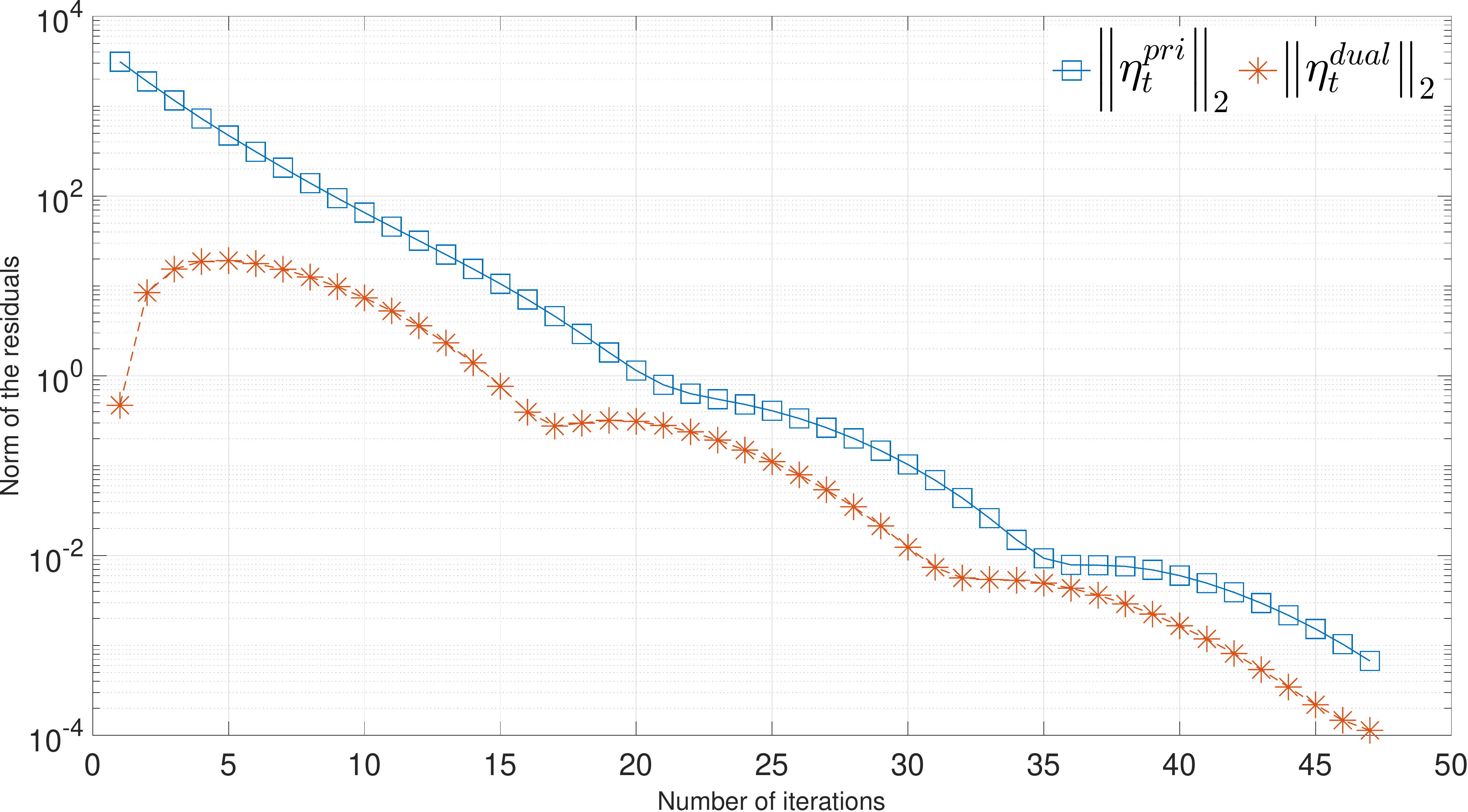}}
\caption{The magnitude of primal residual and dual residual over iterations.}
\label{fig_converg}
\end{figure}

To present the convergence behaviour of the algorithm, Fig.~\ref{fig_converg} shows the values of $\left\| \boldsymbol{\eta }_{t}^{pri} \right\| _2$ and $\left\| \boldsymbol{\eta }_{t}^{dual} \right\| _2$ after different number of iterations for the set up in Fig.~\ref{fig_full}(b). After only 5 iterations, the dual residual began to decrease and local images began to reach consensus.  

\begin{figure}[!htbp]
	\centering
	\subfloat[]{\includegraphics[width=1.74 in]{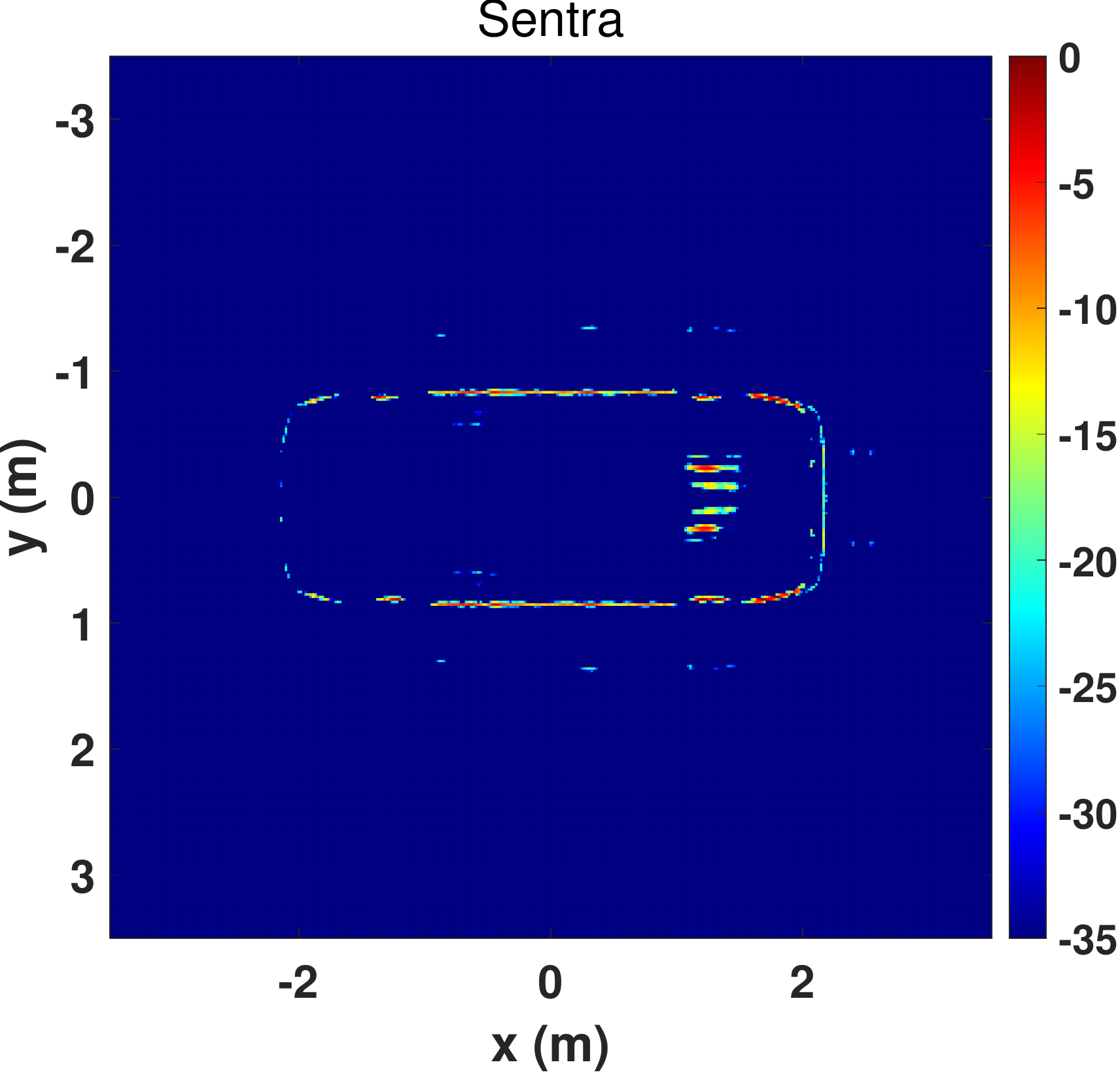}}
	\hfil
	\subfloat[]{\includegraphics[width=1.74 in]{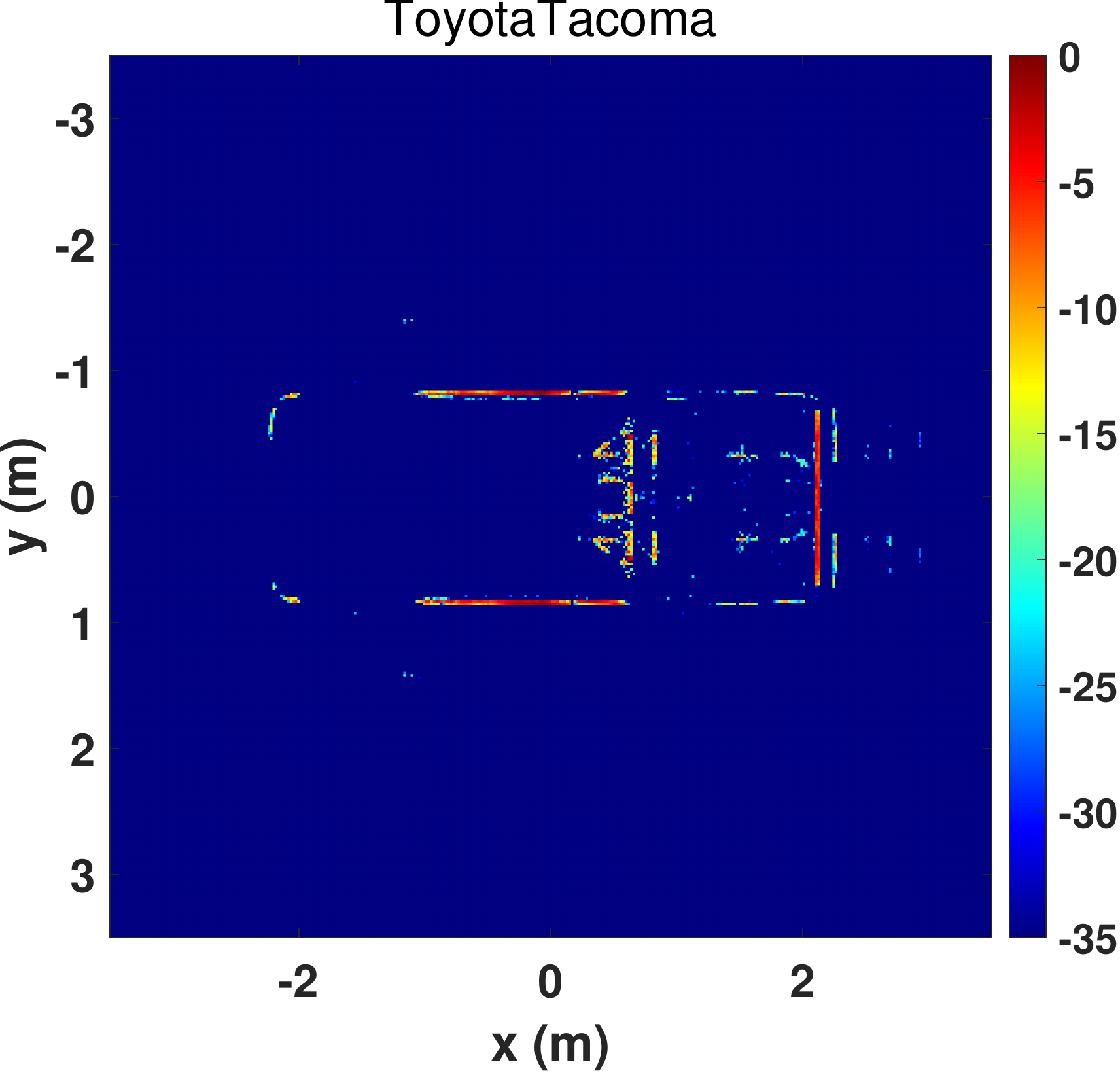}}
	\caption{Full data results of two other models. (a) Sentra; (b) ToyotaTacoma.}
	\label{fig_others}
\end{figure}

However, in Fig.~\ref{fig_full}(b), the left part of the image is missing due to the overlarge weight on the sparsity regularizer. By keeping $\beta = 50$ and increasing $\mu$ to 200 and 300, the targets shown in Fig.~\ref{fig_full}(c) and (d) became more continuous, but the layover artifacts also become slightly stronger. It is a common problem in regularization-based optimization, where hyperparameter tuning is needed for a trade-off between data fidelity and regularization. For the data of two other models, Fig.~\ref{fig_others} shows the full data results with hyperparameters $\mu =100$ and $\beta =50$. The high-quality shown in the obtained images could improve the performance of post-processing like target identification.

\subsection{Results from limited data}
\begin{figure}[!htbp]
	\centering
	\subfloat[]{\includegraphics[width=1.74 in]{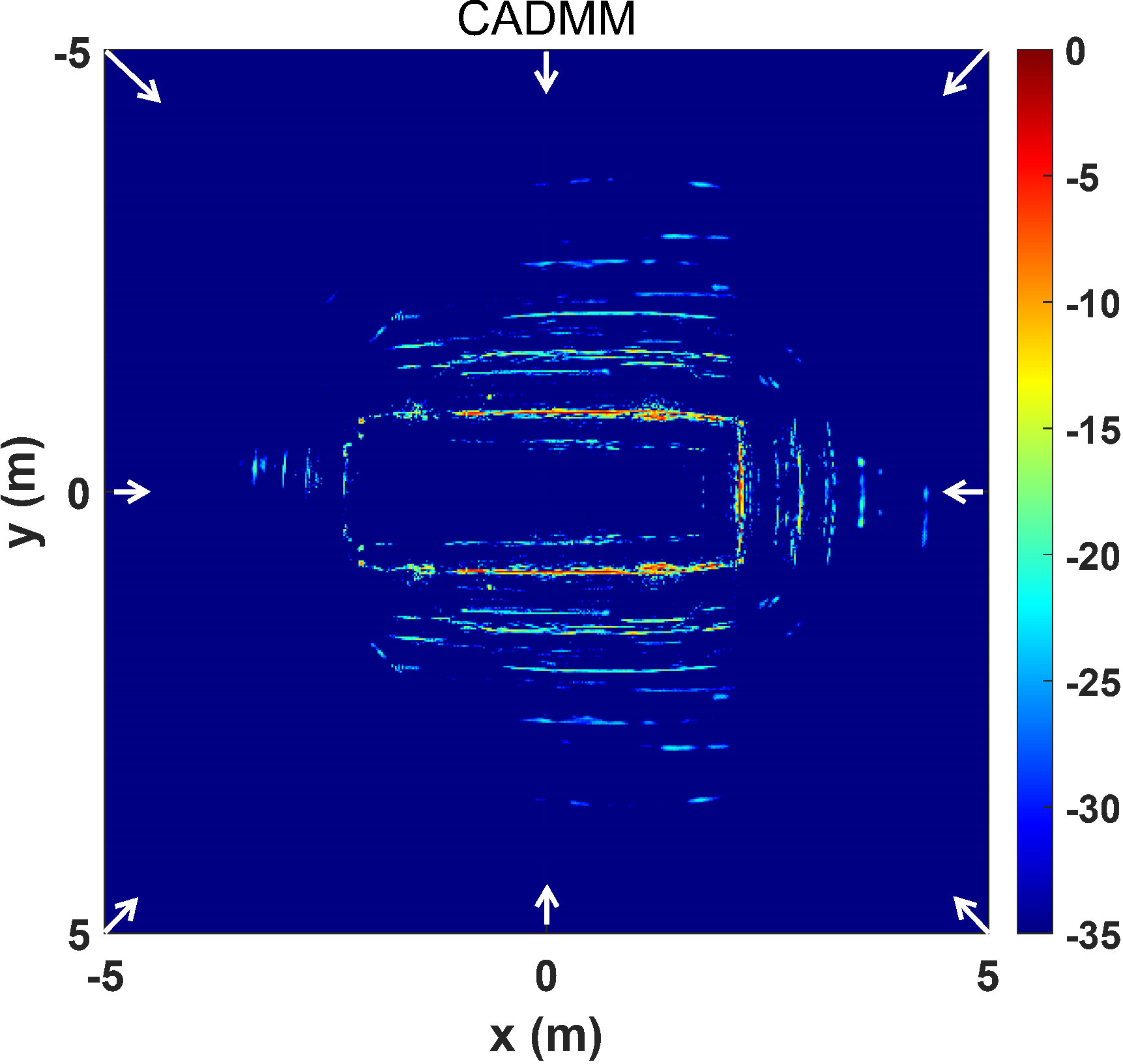}}
	\hfil
	\subfloat[]{\includegraphics[width=1.74 in]{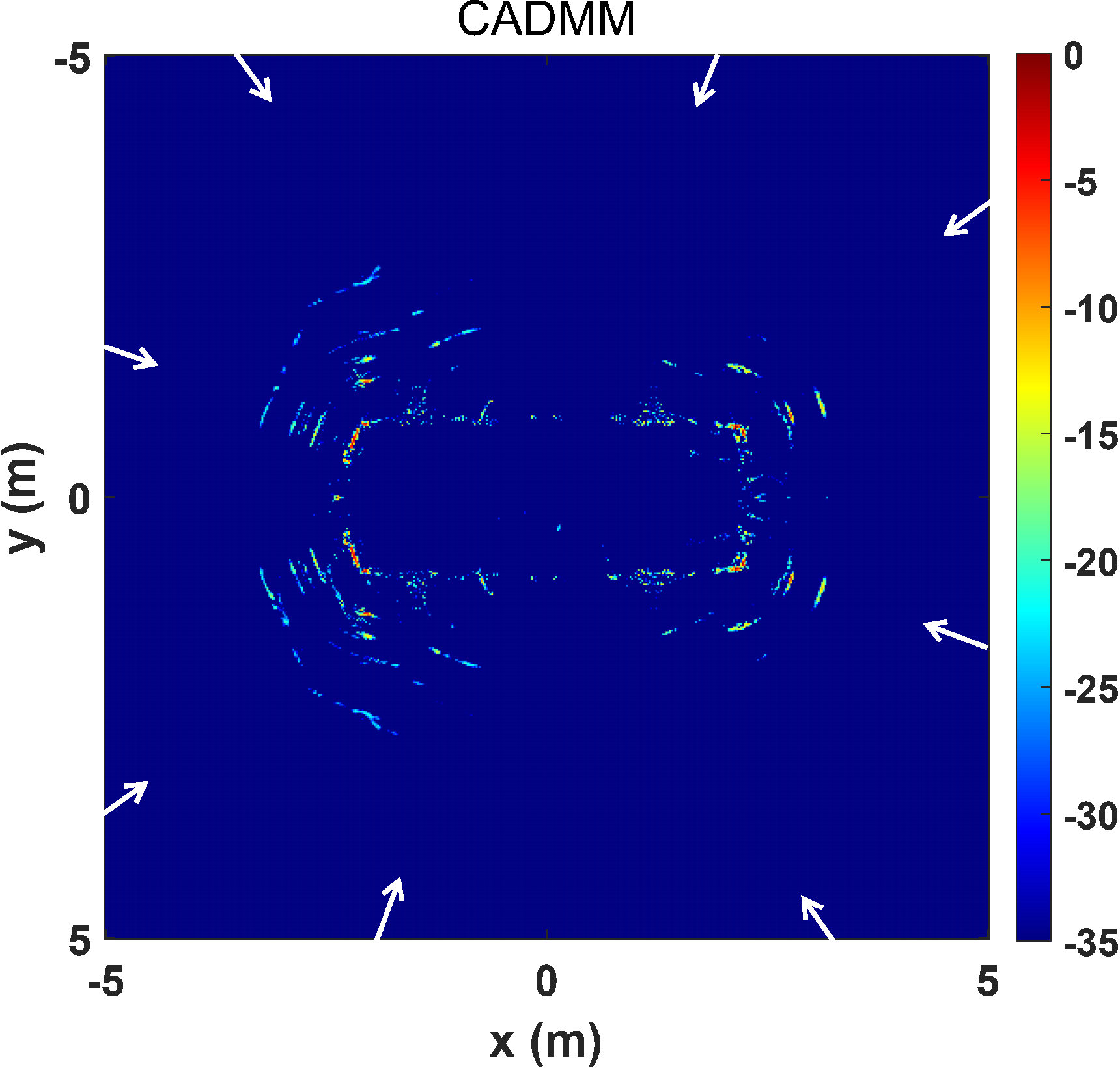}}
	\hfil
	\subfloat[]{\includegraphics[width=1.74 in]{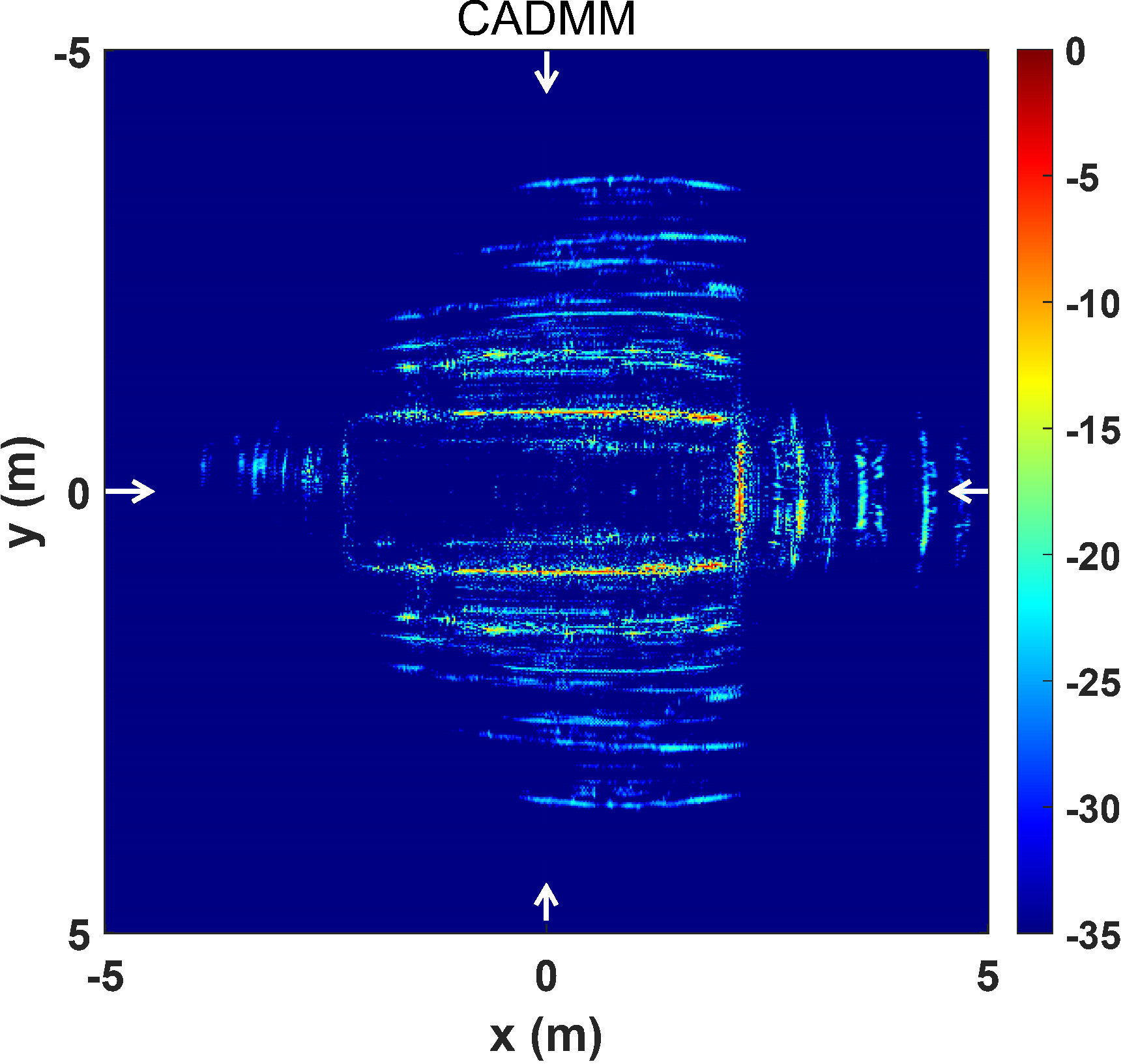}}
	\hfil
	\subfloat[]{\includegraphics[width=1.74 in]{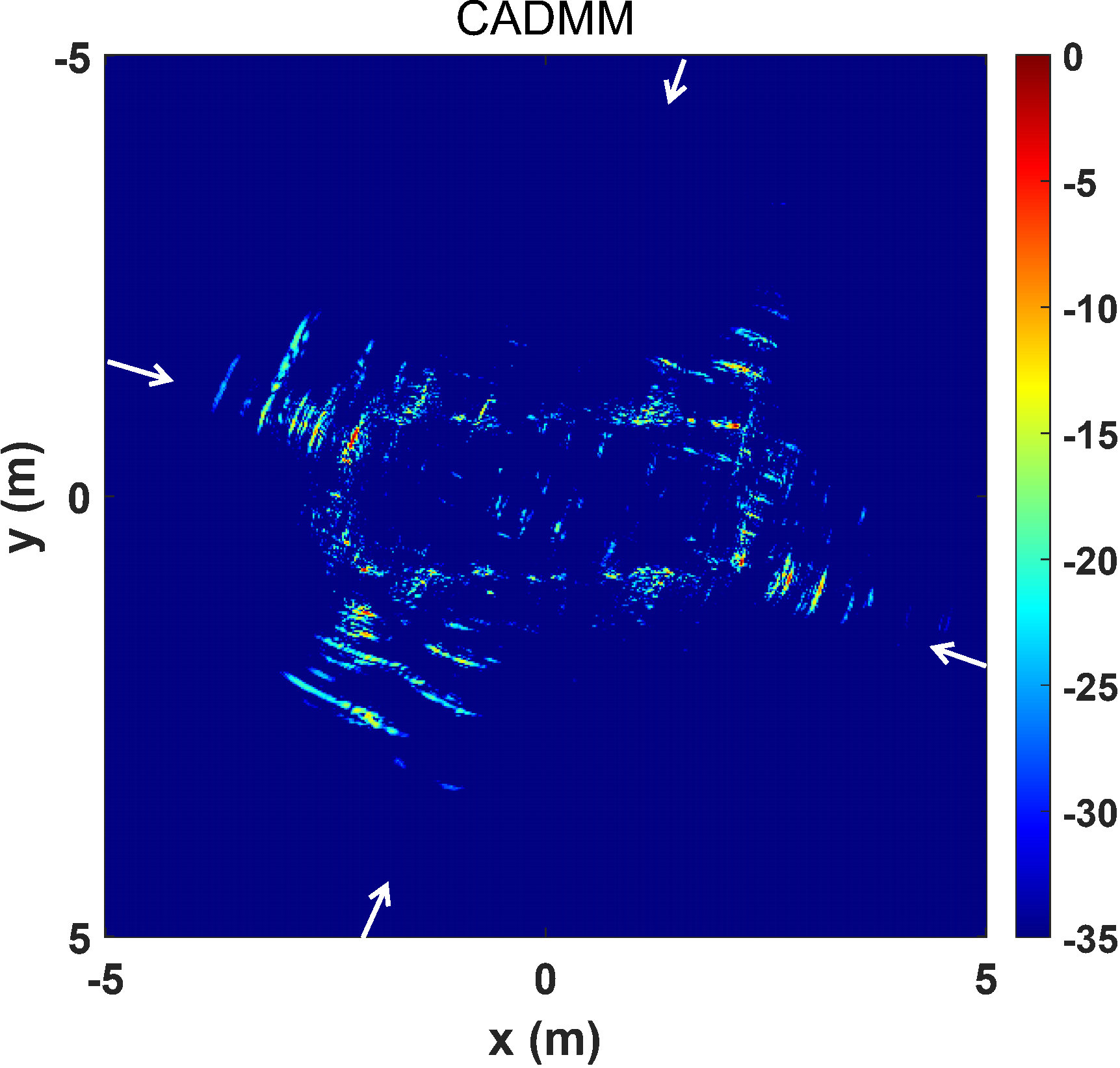}}
	\caption{Limited data results. (a) ${N_\theta } = 8$; (b) ${N_\theta } = 8$ (non-perpendicular); (c) ${N_\theta } = 4$; (d) ${N_\theta } = 4$ (non-perpendicular).}
	\label{fig_sub}
\end{figure}
In practice, it is not likely we could have $360^\degree$ data. Instead, we might only get a limited number of APC clusters with a limited angular extent. For the four elevation angles, by keeping the angular extent of each cluster as $18^\degree$ and reducing the number of clusters in azimuth to ${N_\theta } = 8$ and ${N_\theta } = 4$, CADMM imaging results from the limited data are shown in Fig.~\ref{fig_sub}, where the white arrows indicate the center azimuth direction of APC clusters. It is shown that, the image quality is more inferior when the data and angle diversity is limited. Besides, it also needs more iterations for convergence. As the orientation of targets is arbitrary, for the general cases when the center azimuth angle of APC clusters are not perpendicular to the targets' facades, the reconstructed images (Fig.~\ref{fig_sub}(b) and (d)) suffer from severe deterioration. Improvements can be made from different aspects in the next stages. In terms of architecture, MIMO technique can provide more angular diversities for limited number of distributed antennas. As for the imaging algorithm, random sampling, smoothness, and low-rankness can be exploited to enhance imaging quality.

\section{Conclusion}
In this paper, a new imaging algorithm based on the CADMM framework is proposed to mitigate the artifacts in distributed radar imaging. During the optimization process, because of the additional consensus constraints, the common features of the local images are extracted while the spatially-variant artifacts are suppressed  in a data-driven manner. The measurements from distributed antenna clusters are fused to reach consensus through the proposed algorithm. The final converged global image presents weaker artifacts than composite imaging. The results of simulated data have verified the effectiveness of the proposed method. Further research might include enhancing the imaging performance under limited data and extending 2D imaging to 3D.
\vspace{12pt}
% \color{red}
% IEEE conference templates contain guidance text for composing and formatting conference papers. Please ensure that all template text is removed from your conference paper prior to submission to the conference. Failure to remove the template text from your paper may result in your paper not being published.
\bibliographystyle{IEEEtran}
% \bibliography{strings,bibliography}
\bibliography{bibliography}
\end{document}